\documentclass[graybox]{svmult}

\usepackage{mathptmx}       
\usepackage{helvet}         
\usepackage{courier}        
\usepackage{type1cm}        
\usepackage{makeidx}         
\usepackage{graphicx}        
\usepackage{multicol}        
\usepackage[bottom]{footmisc}

\makeindex             

\usepackage{amsmath}
\usepackage{amsfonts}
\usepackage{tikz}
\usepackage{mathtools}

\usepackage{enumerate}
\usepackage{braket} 

\usepackage{siunitx}

\newcommand{\tauint}{\int_0^\beta \mathrm{d}\tau_1\int_{\tau_1}^\beta \mathrm{d}\tau_2\ldots \int_{\tau_{K-1}}^\beta \mathrm{d}\tau_K}
\newcommand{\intsum}{\ensuremath{\int\hspace{-10pt}\sum}}
\newcommand{\occconfig}[1]{\ensuremath{{ \{ #1 \} }}}
\newcommand{\op}[1]{\ensuremath{\hat{#1}}}

\DeclarePairedDelimiter\abs{\lvert}{\rvert}
\providecommand{\abs}[1]{\ensuremath{\abs{#1}}}
\DeclarePairedDelimiter\mittelwert{\langle}{\rangle}
\providecommand{\mittelwert}[1]{\ensuremath{\mittelwert{#1}}}

\DeclareMathOperator{\Tr}{Tr}

\DeclarePairedDelimiter\klammern{(}{)}
\providecommand{\klammern}[1]{\ensuremath{\klammern{#1}}}
\DeclarePairedDelimiter\kommutator{[}{]}
\DeclarePairedDelimiter\antikommutator{\{}{\}}
\DeclarePairedDelimiter\braces{\{}{\}}
\DeclarePairedDelimiter\brackets{[}{]}
\providecommand{\kommutator}[1]{\ensuremath{\kommutator{#1}}}
\providecommand{\antikommutator}[1]{\ensuremath{\kommutator{#1}}}
\providecommand{\braces}[1]{\ensuremath{\braces{#1}}}

\providecommand{\text}[1]{\ensuremath{\text{#1}}}
\providecommand{\intertext}[1]{\ensuremath{\intertext{#1}}}

\providecommand{\braket}[1]{\ensuremath{\braket{#1}}}
\providecommand{\bra}[1]{\ensuremath{\bra{#1}}}
\providecommand{\ket}[1]{\ensuremath{\ket{#1}}}
\providecommand{\binom}[2]{\binom{#1}{#2}}
\newcommand{\Ham}{\op{H}}
\newcommand{\W}{\op{W}}
\newcommand{\ladderup}{\ensuremath{\op{a}^\dagger}}
\newcommand{\ladderdown}{\ensuremath{\op{a}^{\vphantom{\dagger}}}}

\newcommand{\definedby}{\ensuremath{\coloneqq}}
\newcommand{\defines}{\ensuremath{\eqqcolon}}

\providecommand{\SI}[2]{\SI{#1}{#2}}
\providecommand{\num}[1]{\num{#1}}

\begin{document}

\title*{Introduction to Configuration Path Integral Monte Carlo}
\author{T. Schoof, S. Groth and M. Bonitz}
\institute{T. Schoof \at Kiel University, Department of Physics, Leibnizstr. 15 , Germany; \email{schoof@theo-physik.uni-kiel.de}
\and S. Groth \at Kiel University, Department of Physics, Leibnizstr. 15, Germany; \email{groth@theo-physik.uni-kiel.de}
\and M. Bonitz \at Kiel University, Department of Physics, Leibnizstr. 15, Germany; \email{bonitz@physik.uni-kiel.de}}
%
%
\maketitle

\abstract{In low-temperature high-density plasmas quantum effects of the electrons are becoming increasingly important. This requires the development of new theoretical and computational tools. 
Quantum Monte Carlo methods are among the most successful approaches to first-principle simulations of many-body quantum systems. In this chapter we present a recently developed method---the configuration path integral Monte Carlo (CPIMC) method for moderately coupled, highly degenerate fermions at finite temperatures. It is based on the second quantization representation of the $N$-particle density operator in a basis of (anti-)symmetrized $N$-particle states (configurations of occupation numbers) and allows to tread arbitrary pair interactions in a continuous space.  
\newline\indent
We give a detailed description of the method and discuss the application to electrons or, more generally, Coulomb-interacting fermions. As a test case we consider a few quantum particles in a one-dimensional harmonic trap. Depending on the coupling parameter (ratio of the interaction energy to kinetic energy), the method strongly reduces the sign problem as compared to direct path integral Monte Carlo (DPIMC) simulations in the regime of strong degeneracy which is of particular importance for dense matter in laser plasmas or compact stars. In order to provide a self-contained introduction, the chapter includes a short introduction to Metropolis Monte Carlo methods and the second quantization of quantum mechanics.
}

\section{Introduction}
\index{quantum plasma!simulations|(}
Interacting fermionic many-body systems are of great interest in many areas of physics. These include atoms and molecules, electron gases in solids, partly ionized and dense plasmas in strong laser fields, and astro-physical plasma or the quark-gluon plasma~\cite{paper_1,bonitz_complex_2010,haberland_harmonics_2001,redmer_progress_2010,holst_electronic_2011,bonitz_introduction_2010}. In these systems the quantum behavior of the electrons (and, at high density, also of the ions) plays an important role. Similar relevance of quantum effects is observed in low-temperature plasmas at atmospheric pressure, in particular, when the plasma comes in contact with a solid surface. For the quantum treatment of electrons in the latter case, see the chapter by Heinisch et al.

Thermodynamic properties of these system can only be calculated with huge difficulties if long range interactions as well as degeneracy of the quantum particles become important. 
Many models and approximations exist, including quantum hydrodynamics (QHD), cf. the chapter by Khan et al. in this book, but these are often inaccurate or not controllable. For that reason \emph{first-principle} computer-simulation are of great importance.
Among the most successful methods for the simulation of interacting many-body quantum systems are the density functional theory (DFT)~\cite{tse_ab_2002}, many-body theories like, e.g. Green functions~\cite{kremp_quantum_2005,bonitz_quantum_1998,balzer-book}, and quantum Monte Carlo\index{simulation!quantum Monte Carlo} (QMC) methods. Nevertheless, the \emph{ab initio} simulation of fermions is still an unsolved problem.
\par
The main idea of finite-temperature QMC methods is based on the description of the system in terms of the Feynman path integral~\cite{feynman_quantum_1965}. In this formulation a quantum system in thermodynamic equilibrium can be described by classical paths in an effective ``imaginary time''. While, in principle, exact results can be obtained for arbitrary large particle numbers for bosons~\cite{ceperley_path_1995, filinov_berezinskii-kosterlitz-thouless_2010}, Monte Carlo (MC) methods for fermions suffer from the so-called \emph{fermion sign problem}~\cite{ceperley_buch, troyer_computational_2005}, which leads to exponentially increasing statistical errors in dependence of the particle number. The mathematical reason for this is the alternating sign of the terms that contribute to the expectation values due to the antisymmetry of the fermionic wave function under particle permutations. The physical origin is the Pauli principle. There exist several approaches to reduce or even avoid the sign problem. The \emph{fixed-nodes} path integral Monte Carlo approximation can be considered as one of the most successful methods for highly degenerate systems like warm dense matter~\cite{ceperley_path-integral_1992,driver_all-electron_2012,ferrario_coupled_????,morales_equation_2010}.  It uses the knowledge about the nodal surface structure of a trial density matrix to completely avoid the sign problem. However, the choice of the trial density matrix introduces systematical errors that are difficult to estimate.  Another approach is the multi-level-blocking algorithm~\cite{egger_crossover_1999}. Exact direct path integral Monte Carlo (DPIMC) calculations are also possible with the use of several optimizations ~\cite{filinov_hot-plasma_2001,filinov_wigner_2001} but this is possible only for sufficiently high temperature (moderate	 degeneracy).
\index{quantum plasma!simulations|)}
\par
In this chapter we will present a different approach---the configuration path integral Monte Carlo (CPIMC) method~\cite{schoof_configuration_2011}. The main idea of this approach is to evaluate the path integral not in configuration space (as in DPIMC), but in second quantization. This leads to paths in Slater determinant space in occupation number representation instead of paths in coordinate space. The idea is based on the continuous time path integral method~\cite{prokofev_exact_1996, prokofev_exact_1998}, which is widely used for lattice models such as the Hubbard model, see~\cite{troyer_non-local_2003}, and for impurity models, see~\cite{gull_continuous-time_2011}. These models are described by simplified Hamiltonians where the interaction is typically of short-range. However, for many systems, in particular plasmas, the long-range Coulomb interaction is of central importance. In this chapter we will show how the CPIMC idea can be used to calculate thermodynamic properties of plasmas or, more generally, spatially continuous fermionic systems with arbitrary pair interactions. 
\par
The chapter will start with a short introduction to Metropolis Monte Carlo methods and an advanced description of the estimation of statistical errors. We will then introduce the formalism of the second quantization as far it is needed to understand the presented algorithm and give a short overview over the equations of statistical quantum physics that describe the thermodynamics of the systems of interest in equilibrium.  In the main part we will derive the underlying formulas of CPIMC in detail and give a complete description of the MC algorithm. At the end of the chapter a demonstrative application of the method to a one-dimensional system of Coulomb-interacting fermions in a harmonic trap will be discussed.
\section{Monte Carlo\label{Monte_Carlo_chap}}
\index{simulation!Monte Carlo}
In the first section of this chapter, the \emph{Metropolis algorithm}\index{Metropolis algorithm}, which has been developed by Metropolis \emph{et. al.} in 1953 \cite{metropolis_equation_1953}, shall be explained in a fairly general form. The next section, where we will briefly illustrate how to estimate the error of quantities computed with the \emph{Metropolis algorithm}, is based on \cite{wolfhard_janke_statistical_2009}. We mention that similar Monte Carlo methods are used also in the chapters by Thomsen et al. (classical Metropolis MC) and Rosenthal et al. (kinetic MC) of this book. 
\subsection{Metropolis algorithm}
Our ultimate goal is the computation of thermodynamic properties of quantum many-body systems in equilibrium. However, the following considerations are completely general and are valid for classical and quantum systems. In the classical case, we only have to remove the operator heats of the observables. Regardless of the chosen ensemble, expectation values of observables $\op{O}$  are always of the form
\begin{align}
\braket{\op{O}}=\intsum_{\;C} O(C)\frac{W(C)}{Z}\qquad\text{with}\qquad Z=\intsum_{\;C} W(C)\;.
\label{general_observable}
\end{align}    
The partition function\index{partition function} $Z$ is given by the sum over all weights $W(C)$, where the multi-variable $C$ defines exactly one contribution to the partition function, i.e. one specific weight. Moreover, we refer to the multi-variable $C$ as a configuration or a microstate of the system. Such a configuration is generally defined by a set of discrete and/or continuous variables, which is why the symbol $\displaystyle\intsum_C$ was introduced in (\ref{general_observable}) denoting the integration or summation over all these variables. Further, $O(C)$ denotes the value of the observable in the configuration $C$. We can interpret $P(C)=\frac{W(C)}{Z}$ as the probability of the system to be in the configuration $C$ if and only if the weights $W(C)$ are always real and positive. Hence, the expectation value of an observable [Eq. (\ref{general_observable})] is given by the sum over all possible values of the observable $O(C)$ weighted with its corresponding probability. \par 
The difference between classical and quantum systems as well as the choice of the ensemble (microcanonical, canonical or grand canonical ensemble) enter in the actual form of the weights and in the set of variables necessary to define a configuration. [In Chap.~1 by Thomsen et al., Metropolis Monte Carlo of a classical system in the canonical ensemble is described.] For a quantum  system, the partition function is given by the trace over the $N-$particle density operator $\op{\rho}$ discussed in detail in Sec. \ref{density_op_chap} below:
\begin{align}
Z = \Tr{\op{\rho}}\;.
\end{align}
Each choice of a different basis in which the trace is performed leads to a different Monte Carlo algorithm. For example, performing the trace in coordinate space results in the DPIMC method \cite{ceperley_path_1995}, whereas the occupation number representation yields the CPIMC method described in Sec. \ref{CPIMC_Chap} and \cite{schoof_configuration_2011}.\par    

Now we will continue with  the general consideration of the \emph{Metropolis algorithm} for quantum and classical systems. 
Using the \emph{Metropolis algorithm}, one can generate a sequence of configurations $C_i$ which are distributed with probability $P(C_i)$. However, $P(C_i)$ involves the partition function $Z$ (normalization of the distribution), that is typically very hard to compute. 
To avoid this problem, one defines a transition probability $v(C_i\to C_{i+1})$ for the system that characterizes the transition into the configuration $C_{i+1}$ starting from $C_i$. To ensure that the configurations $C_i$ are distributed with $P(C_i)$, the transition probability has to fulfill the \emph{detailed balance condition}\index{detailed balance condition}\footnote{In fact, there are weaker conditions on the transition probabilities, but in practice, one uses the \emph{detailed balance}.  }:
\begin{align}
P(C_i)v(C_i\to C_{i+1})=P(C_{i+1})v(C_{i+1}\to C_i)\;.
\label{detailed_balance}
\end{align}  
A possible solution of this equation is given by
\begin{align}
v(C_i\to C_{i+1})=\min{\left[1,\frac{P(C_{i+1})}{P(C_i)}\right]}=\min{\left[1,\frac{W(C_{i+1})}{W(C_i)}\right]}\;.
\label{transitionprob}
\end{align}
Now the algorithm works as follows. Suppose the system is in the configuration $C_i$. A Monte Carlo step then consists of proposing a certain change of the configuration converting it into $C_{i+1}$. After that, a random number from $[0,1)$ is generated, and the transition probability is calculated according to Eq. (\ref{transitionprob}). If the random number is smaller than $v(C_i\to C_{i+1})$, then the change (Monte Carlo step) is accepted. Otherwise, the system stays in the previous configuration, i.e. it is $C_{i+1}=C_{i}$. Starting from an arbitrary configuration $C_0$, after some steps, the configurations will be distributed with $P(C)$ forming a so-called \emph{Markov chain}\index{Markov chain}. We refer to the number of steps which are necessary for the initial correlations to  vanish as the \emph{equilibration time}.
\par In addition to the \emph{detailed balance}, we have to make sure the transition probabilities are ergodic, i.e. it must be possible to reach every configuration with $W(C)>0$ in a finite number of steps.
 In practice, one will usually need a few different Monte Carlo steps to ensure the ergodicity. These steps address different degrees of freedom of the configurations, which, of course, depend on the actual form of the partition function, Eq. (\ref{general_observable}). Furthermore, in the majority of cases, it will be more efficient not to propose every change of the configuration with equal probability. Therefore, we split the transition probability into a sampling probability $T(C_i\to C_{i+1})$ and an acceptance probability $A(C_i\to C_{i+1})$ for that specific change, i.e. it is
\begin{align}
v(C_i\to C_{i+1})=A(C_i\to C_{i+1})T(C_i\to C_{i+1})\;.
\end{align} 
Inserting this factorization into the \emph{detailed balance}, one readily obtains the solution for the acceptance probability
\begin{align}
A(C_i\to C_{i+1})=\min{\left[1,\frac{T(C_{i+1}\to C_i)W(C_{i+1})}{T(C_{i}\to C_{i+1)}W(C_{i})}\right]}\label{Akz}\;. 
\end{align} 
This generalization is also called the \emph{Metropolis Hastings algorithm}.
The algorithm is most efficient if the acceptance probability is about 50\%. A suitable choice of the sampling probability can help optimizing the acceptance probability.\par
Once we have generated a \emph{Markov chain} of length $N_{MC}$ via the \emph{Metropolis algorithm}, a good estimator for the expectation value of an observable is given by
\begin{align}
\braket{\op{O}}\approx\frac{1}{N_{MC}}\sum_{i=1}^{N_{MC}}O(C_i)\label{schaetz}=:\overline{O}\;, 
\end{align}
which directly follows from Eq. (\ref{general_observable}) and the fact that all configurations $C_i$ in the \emph{Markov chain} are distributed with $P(C_i)$.\par
Unfortunately, the weight $W(C)$ is not always strictly positive, so we can not always interpret $\frac{W(C)}{Z}$ as a probability. Especially for systems of fermions, there are many sources of sign changes in the weights. However, we can still apply the \emph{Metropolis algorithm} if we define a new partition function
\begin{align}
Z^\prime:=\sum_C|W(C)|\;,
\label{new_system}
\end{align}
and rewrite the expectation values as follows:
\begin{align}
\braket{\op{O}}=\frac{\sum_C O(C)W(C)}{\sum_CW(C)}=\frac{\frac{1}{Z^\prime}\sum_{C}O(C)S(C)|W(C)|}{\frac{1}{Z^\prime}\sum_CS(C)|W(C)|}
=\frac{\braket{\op{O}S}^\prime}{\braket{S}^\prime}\label{sign}\;,
\end{align}     
where $S(C)=\text{sgn}[W(C)]$ denotes the sign of the weight. Obviously, the expectation value of $\op{O}$ in the physical system, described by the partition function $Z$, is equivalent to the expectation value of $\op{O}S$ divided by the expectation value of the sign (\emph{average sign}), both averaged in the primed system described by $Z'$, cf. Eq.~(\ref{new_system}). For that system, we can generate a \emph{Markov chain} if we simply insert the modulus of the weights $\abs{W(C)}$ in Eq. (\ref{Akz}) when computing the acceptance probability. Combining Eq. (\ref{schaetz}) and (\ref{sign}), we obtain an expression for the estimator of the observable in the real system:
\begin{align}
\braket{\op{O}}\approx\frac{\sum_{i=1}^{N_{MC}} O(C_i)S(C_i)}{\sum_{i=01}^{N_{MC}}S(C_i)}=:\frac{\overline{OS}^{\;\prime}}{\overline{S}^{\;\prime}}\;=\overline{O}.
\label{estimator_with_sign}
\end{align}
 Thus, we actually simulate the primed system and calculate quantities of the physical system via Eq. (\ref{estimator_with_sign}).

 \subsection{Error in the Monte Carlo simulation\label{sign_prob_chap}}
 Estimating the errors\index{error!statistical} of quantities computed with the \emph{Metropolis algorithm} is not a trivial task. First, we consider the case of strictly positive weights, where the estimator of an observable is given by Eq. (\ref{schaetz}).
It is reasonable to interpret the values of the observable in each configuration of the \emph{Markov chain} as a series of measurements. The estimator $\overline{O}$, which is obtained by averaging over all measurements, fluctuates statistically around the true expectation value $\braket{\op{O}}$. The error of the arithmetic mean of a measurement series of length $N_{MC}$ is defined by
\begin{align}
\Delta \overline{O}=\frac{\sigma_{\overline{O}}}{\sqrt{N_{MC}}}\;,
\end{align} 
where $\sigma_{\overline{O}}$ is the standard deviation which, for uncorrelated measurements, can be estimated by
\begin{align}
\sigma_{\overline{O}}=\sqrt{\frac{1}{(N_{MC}-1)}\sum_{i=0}^{N_{MC}}(O_i-\overline{O})^2\,,}\qquad\text{with}\qquad O_i=O(C_i).
\end{align}
Certainly, the ``measurement'' series obtained from the \emph{Markov chain} is correlated, since we generate each configuration from the previous one. The correlation of these configurations can be measured by the integrated auto-correlation time
\begin{align}
\tau_{\text{int},O}&=\frac{1}{2}+\sum_{k=1}^{N_{MC}}\frac{\overline{O_iO_{i+k}}-\overline{O}^2}{\sigma_{\bar{O}}^2},\\
\qquad\text{with}\qquad \overline{O_iO_{i+k}}&=\frac{1}{N_{MC}}\sum_{i=1}^{N_{MC}}O_iO_{i+k} \nonumber\;.
\end{align}  
The auto-correlation of the measurements enhances the statistical error by a factor of $\sqrt{2\tau_{\text{int},O}}$, i.e. it is
\begin{align}
\Delta\overline{O}_{\text{auto}}=\Delta\overline{O}\sqrt{2\tau_{\text{int},O}}=\frac{\sigma_{\bar{O}}}{\sqrt{N_{MC}}}\sqrt{2\tau_{\text{int},O}}=\frac{\sigma_{\bar{O}}}{\sqrt{N_{\text{eff}}}}.
\label{auto_error}
\end{align}
Therefore, if we have generated a \emph{Markov-chain} of length $N_{MC}$, we have effectively gained only $\frac{N_{MC}}{2\tau_{\text{int},O}}$ uncorrelated measurements. To keep the correlation small, efficient Monte Carlo steps are essential, which change many degrees of freedom of the configurations simultaneously while still having an acceptance ratio of about 50\%. In practice, acceptance ratios are much smaller than 50\%, and one therefore performs a certain number of Monte Carlo steps (a \emph{cycle}) before again measuring the observables.\par
Things become more complex if the weights are not strictly positive. In that case, we have expressed quantities of the physical system by quantities of the primed system, Eq. (\ref{estimator_with_sign}).   
We simulate the primed system and calculate the r.h.s. of Eq. (\ref{estimator_with_sign}) to obtain $\overline{O}$. The measurements of the quantities $\overline{OS}^{\;\prime}$ and $\overline{S}^{\;\prime}$ are not only auto-correlated but also cross-correlated with each other. The relative statistical error can be estimated according to (cf. \cite{wolfhard_janke_statistical_2009})
 \begin{align}
\frac{\Delta\overline{O}_{\text{auto,cross}}}{\overline{O}}&=\sqrt{\left(\frac{\Delta\overline{OS}^{\;\prime}}{\overline{OS}^{\;\prime}}\right)^2+\left(\frac{\Delta\overline{S}^{\;\prime}}{\overline{S}^{\;\prime}}\right)^2
-\frac{2}{N_{MC}}\frac{\overline{OSS}^{\;\prime}-\overline{OS}^{\;\prime}\;\overline{S}^{\;\prime}}{\overline{OS}^{\;\prime}\;\overline{S}^{\;\prime}}2\tau_{int,OS,S}}\;,
\label{error_root}
\end{align}
 where we have the statistical errors of $\overline{OS}^{\;\prime}$ and $\overline{S}^{\;\prime}$
 \begin{align}
\Delta\overline{OS}^{\;\prime}&=\sqrt{\frac{\overline{(OS)^2}^{\;\prime}-\overline{(OS)}^{\;\prime^{\;2}}}{N_{MC}}2\tau_{\text{int},OS}}\;,\nonumber\\
\Delta\overline{S}^{\;\prime}&=\sqrt{\frac{\overline{S^2}^{\;\prime}-\overline{S}^{\;\prime^{\;2}}}{N_{MC}}2\tau_{\text{int},S}}\;, 
 \end{align}
 each enhanced by their individual auto-correlation time and in the last term the integrated cross-correlation time
 \begin{align}
 \tau_{\text{int},OS,S}=\frac{1}{2}+\sum_{k=1}^{N_{MC}}\frac{\overline{O_iS_iS_{i+k}}^{\;\prime}-\overline{OS}^{\;\prime}\;\overline{S}^{\;\prime}}{\overline{OSS}^{\;\prime}-\overline{OS}^{\;\prime}\;\overline{S}^{\;\prime}}\;,\quad\text{with}\quad
 \overline{O_iS_iS_{i+k}}^{\;\prime}=\frac{1}{N_{MC}}\sum_{i=1}^{N_{MC}}O_iS_iS_{i+k}\;.\nonumber 
 \end{align}
 Assuming we are working in the canonical ensemble, then, from the definition of the partition function and the \emph{average sign}, cf. Eq. (\ref{sign}),  it immediately follows that
  \begin{align}
 \overline{S}^{\;\prime}\approx\braket{S}^\prime=\frac{Z}{Z^\prime}=e^{-\beta N(f-f^\prime)}\leq 1\;,
 \label{sign_decrease}
 \end{align}   
 where $\beta$ is the inverse temperature, $N$ the particle number, and $f^{(\prime)}$ the free energy per particle in the physical (primed) system. Combining Eqs.~(\ref{error_root}) and (\ref{sign_decrease}) yields
 \begin{align}
\frac{\Delta \overline{O}_{\text{auto,sign}}}{\overline{O}}\propto \frac{1}{\overline{S}^{\;\prime}\sqrt{N_{MC}}}\approx \frac{1}{\sqrt{N_{MC}}}e^{\beta N(f-f^\prime)}\;.
\label{eq:mc_error} 
 \end{align}
 The relative error of an observable grows exponentially with the product of the particle number, the inverse temperature and the difference of the free energy per particle in the physical and the primed system. Unfortunately, the error can only be reduced with the square root of the number of Monte Carlo samples (measurements).
 Therefore, the value of the \emph{average sign} of a given system determines if one can compute reliable quantities via the \emph{Metropolis algorithm} for that system.  This severe limitation is called the (fermion) \emph{sign problem}. In practice, an average sign of the order $\approx 10^{-3}$ is the limit for reliable Monte Carlo simulations. Also, one should be aware that even the sign itself has a statistical error of the form (\ref{auto_error}). Besides, we note that the proportionality (\ref{sign_decrease}) can be obtained by a simple Gaussian error propagation of $\overline{O}$, cf. Eq. (\ref{estimator_with_sign}).\par
 The \emph{sign problem} is strongly dependent on the actual representation of the partition function, cf. Eq. (\ref{general_observable}). For some special systems, the \emph{sign problem} can be circumvented by a specific choice of the representation \cite{nakamura_vanishing_1998, lyubartsev_simulation_2005}, but a general solution is highly unlikely, since it has been shown to be NP-complete \cite{troyer_computational_2005}.     
\section{Second quantization}
Second quantization refers to the introduction of creation and annihilation operators for the description of quantum many-particle systems. Thereby, not only the observables are represented by operators but also the wave functions, in contrast to the standard formulation of quantum mechanics (correspondingly called ``first'' quantization), where only the observables are represented by operators. The proper symmetry properties of the bosonic (fermionic) many-particle states are automatically included in the (an\-ti-)commutation relations of the creation and annihilation operators. Simultaneously, also observables can be expressed by these operators. This unifying picture of quantum mechanics is often advantageous when it comes to the description of many-particle systems.   
    
Since CPIMC makes extensive use of the second quantization of quantum mechanics, a brief introduction will be given in this section, where we will focus on those aspects which are necessary for the comprehension of the method. Especially the derivation of the crucial \emph{Slater-Condon rules} shall be outlined. For a detailed introduction to the second quantization see e.g. \cite{helgaker_molecular_2000}. 
\subsection{(Anti-)Symmetric many-particle states}
We consider a system of $N$ identical, ideal (i.e. not interacting) quantum particles which is described by the Hamiltonian $\op{H}_0$. For an ideal system, the Hamiltonian can be written as a sum of one-particle Hamiltonians:
\begin{align}
\Ham_0 &= \sum_{\alpha=1}^N \op{h}_\alpha \label{eq:IdealerHamilton},\\
\op{h}_\alpha &= \frac{\op{p}_\alpha^2}{2m}+\op{v}_\alpha\;,
\end{align}
where $\op{p}_\alpha$ denotes the momentum operator and $\op{v}_\alpha$ the operator of the potential energy of the particle $\alpha$ in an external field. Thus, the subscript $\alpha$ indicates that those operators are acting on states from the one-particle Hilbert space $\mathcal{H}_\alpha$ of the particle $\alpha$.  
We imply that the solutions of the eigenvalue problem
\begin{align}
\op{h}\ket{i} = \epsilon_i \ket{i}\;, \quad \text{with} \quad i\in\mathrm{N}_0
\label{idealbases}
\end{align}
are known. The one-particle states $\ket{i}$ form a complete orthonormal system (CONS) of the one-particle Hilbert space. Further, we assume the states are arranged according to their one-particle energy, i.e. $\epsilon_i\leq\epsilon_j$ for all $i\leq j$. In general, the one-particle states are spin orbitals\index{orbital}, i.e. $\ket{i}\in\mathcal{H}_\text{coord}\otimes\mathcal{H}_\text{spin}$ with the tensor product of the coordinate and spin Hilbert space. Hence, the wave function\index{wave function} $\braket{\vec{r}\sigma|i}=(\bra{\vec{r}\,}\bra{\sigma})\ket{i}=\phi_i(\vec{r},\sigma)$ depends on both the coordinate $\vec{r}$ and the spin projection $\sigma$ of the particle. \par
In the case of an ideal system, the solution of the $N-$particle eigenvalue problem
\begin{align}
\op{H}_0\ket{\Psi}=E\ket{\Psi}
\end{align}
can be constructed from products of one-particle states:
\begin{align}
\ket{i_1i_2\ldots i_N}=\ket{i_1}_1\ket{i_2}_2\ldots\ket{i_N}_N\;. \label{eq:Produktzustand}
\end{align}
The many-body Hilbert space of $N$ particles $\mathcal{H}^N=\bigotimes_{\alpha=1}^N \mathcal{H}_\alpha$ is thus given by the tensor product of one-particle Hilbert spaces. The states $\ket{i_1i_2\ldots i_N}$ form a basis of $\mathcal{H}^N$, and the used notation indicates that particle $\alpha$ is in the state $\ket{i_\alpha}$. \par
Apart from some special exceptions as e.g. Axions, only totally symmetric or antisymmetric states with respect to arbitrary two-particle exchanges\index{exchange} are physically realized, i.e. states with 
\begin{align}
\ket{\ldots i_{\alpha} \ldots i_{\beta}\ldots}_\pm = \pm \ket{\ldots i_\beta \ldots i_\alpha \ldots} \quad \forall \alpha,\beta\;.
\end{align} 
Particles that are described by symmetric states (upper sign) are called bosons and those with antisymmetric states (lower sign) fermions\index{fermion}. The reason for this symmetry of $N-$particle states lies in the indistinguishability of the quantum particles, whereby physical properties of the system cannot change under particle exchange. The spin-statistics-theorem states that fermions have half-integer and bosons integer spin. \par
Totally (anti-)symmetric states can be constructed from the product states (\ref{eq:Produktzustand}) by summing up all $N!$ permutations of $N$ particles:
\begin{align}
\ket{i_1i_2\ldots i_N}_\pm = \frac{1}{\mathcal{N}} \sum_{\mathcal{P}\in S_N}(\pm 1)^P \op{\mathcal{P}} \ket{i_1}_1\ket{i_2}_2\ldots\ket{i_N}_N\;, \label{eq:Antisymmetrisierung}
\end{align}  
where $\op{\mathcal{P}}$ is the $N-$particle permutation operator that can be constructed from a composition of two-particle exchanges. The normalization factor $\mathcal{N}$ is given by
\begin{align}
\mathcal{N}=\begin{cases}
\sqrt{N!\prod_{i=0}^\infty n_i}\;, \quad &\text{for bosons}, \\[2ex]
\sqrt{N!}\;,\quad &\text{for fermions},
\end{cases}\;
\end{align}
where $n_i$ denotes the number of one-particle states $\ket{i}$ in the product state. The hermitian operator $\op{S}(\op{A})$
\begin{align}
\op{S}/\op{A}=\frac{1}{\mathcal{N}} \sum_{\mathcal{P}\in S_N}(\pm 1)^P \op{\mathcal{P}}\,,
\end{align}
is referred to as the (anti-)symmetrization\index{anti-symmetrization} operator. For fermions, the sign of each summand is determined by the number of two-particle exchanges $P$ in the permutation operator $\op{\mathcal{P}}$. Further, arbitrary (anti-)symmetric states can be constructed from linear combinations of these (anti-)symmetric states, Eq. (\ref{eq:Antisymmetrisierung}). Therefore, they form a basis in the (anti-)symmetric Hilbert space $\mathcal{H}_\pm^N\subset \mathcal{H}^N$. The operator $({\mathcal{N}}/{N!})\op{S}$ or \ $({\mathcal{N}}/{N!})\op{A}$ is a projection operator on the respective subspace, it is $\op{S}^2=({N!}/{\mathcal{N}})\op{S}$ and\ $\op{A}^2=({N!}/{\mathcal{N}})\op{A}$, respectively and also $\op{S}\op{A}=0$. 

In coordinate-spin representation, the anti-symmetric product states of fermions can be written as determinants (called \emph{Slater determinants}):
\begin{align}
\Psi(x_1,\ldots,x_N)_-=\frac{1}{N!}\begin{vmatrix}
\phi_{i_1}(x_1) & \phi_{i_2}(x_1) & \cdots & \phi_{i_N}(x_1) \\
\phi_{i_1}(x_2) & \phi_{i_2}(x_2) & \cdots & \phi_{i_N}(x_2) \\
\vdots & \vdots & & \vdots \\
\phi_{i_1}(x_N) & \phi_{i_2}(x_N) & \cdots & \phi_{i_N}(x_N) \\
\end{vmatrix}. \label{eq:SlaterDeterminante}
\end{align}    
For bosons, one obtains instead permanents. From this representation of antisymmetric states it becomes obvious that there is none with a one-particle state occurring twice in the product state, for a determinant with two equal columns vanishes. This is also known as the famous \emph{Pauli principle}. For bosons, there is no such restriction.   

\subsection{Occupation number representation}
Due to the (anti-)symmetrization of the many-particle states (\ref{eq:Antisymmetrisierung}), such states are entirely characterized by the occurrence frequency of each one-particle state in the product state. The number of particles $n_i$ in the one-particle state $\ket{i}$ is called the occupation number\index{occupation number} (of the $i$-th state/orbital). Since the complete set of occupation numbers, denoted with $\occconfig{n}$, defines an (anti-)symmetric
many-particle state, we can also write
\begin{align}
\ket{i_1\ldots i_N}_\pm \equiv \ket{n_0 n_1 n_2 \ldots} \defines \ket{\occconfig{n}}\;, \label{eq:Besetzungszahlvektor}
\end{align}
with 
\begin{align}
n_i\in \begin{cases}
\mathbb{N}_0 &,\quad \text{for bosons} \nonumber\\
\{0,1\} &,\quad \text{for fermions}
\end{cases}\;.
\end{align}
The order of the occupation numbers equals the order of the one-particle states in the product state, which can be arbitrary but must remain fixed for further calculations. The \emph{Pauli principle}\index{Pauli principle} for fermions is automatically included in the restriction on the occupation numbers to be either zero or one.\par
If we only consider many-particle states with fixed particle number, $N=\sum_{i=0}^\infty n_i$, then the states $\ket{\occconfig{n}}$ form a CONS of the (anti-)symmetric $N-$particle Hilbert space $\mathcal{N}^N_{\pm}$ with the 
 orthogonality relation 
 \begin{align}
\braket{\occconfig{n}|\occconfig{\bar{n}}}=\prod_{i=0}^\infty \delta_{n_i,\bar{n}_i}\defines \delta_{\occconfig{n},\occconfig{\bar{n}}}\,,\label{eq:FockOrthogonal}
\end{align}
and the completeness relation
\begin{align}
\sum_{\occconfig{n}}\ket{\occconfig{n}}\bra{\occconfig{n}}\delta_{\sum_i n_i,N}=\op{1}_N\;.\label{eq:HilbertVollständig}
\end{align}
To shorten the notation, we introduced the following abbreviation:
\begin{align}
\sum_{\occconfig{n}}:=\begin{cases}
\sum_{n_0=0}^\infty \sum_{n_1=0}^\infty \dotsm, &\quad \text{for bosons}\\[2ex]
\sum_{n_0=0}^1 \sum_{n_1=0}^1 \dotsm, &\quad \text{for fermions}\\
\end{cases}\;.
\end{align}  
In practical simulations, one has to work with a finite number of one-particle orbitals $N_B$. This equals the approximation  
\begin{align}
\sum_{\occconfig{n}}\approx\begin{cases}
\sum_{n_0=0}^\infty \sum_{n_1=0}^\infty \dotsm\sum_{n_{N_B}=0}^\infty,  &\quad \text{for bosons}\\[2ex]
\sum_{n_0=0}^1 \sum_{n_1=0}^1 \dotsm\sum_{n_{N_B}=0}^1, &\quad \text{for fermions}\\
\end{cases}\;.
\end{align}  
Given a certain one-particle basis, each antisymmetric state $\ket{\occconfig{n}}$ can be uniquely identified with a \emph{Slater determinant} (\ref{eq:SlaterDeterminante}). The relations (\ref{eq:FockOrthogonal}) and (\ref{eq:HilbertVollständig}) are consistent with the respective relations for determinants. For that reason, it is common to also refer to the states $\ket{\occconfig{n}}$ as \emph{Slater determinants}. \par
Now we drop the restriction of a fixed number of particles. The inner product in Eq. (\ref{eq:FockOrthogonal}) is still well defined for states with different particle numbers; only the completeness relation is slightly  modified
\begin{align}
\sum_{\occconfig{n}}\ket{\occconfig{n}}\bra{\occconfig{n}}=\op{1}\;.\label{eq:FockVollständig}
\end{align}
The states $\ket{\occconfig{n}}$ thereby form a CONS in the so-called Fock space $\mathcal{F}_{\pm}=\mathcal{H}^0\oplus\mathcal{H}^1\oplus\mathcal{H}^2_{\pm}\ldots$, which contains (anti-)symmetric states of varying particle number. 
Consequently, any state of the Fock space can be written as a linear combination of the \emph{Slater determinants} $\ket{\occconfig{n}}$:
\begin{align}
\ket{\Psi}=\sum_{\occconfig{n}}c_{\occconfig{n}} \ket{\occconfig{n}}\;.
\label{superposition}
\end{align}
For example, if we consider a general Hamiltonian of interacting particles
\begin{align}
\op{H}=\op{H}_0+\op{W}\;,
\end{align} 
where we added the interaction operator $\op{W}$ to the ideal Hamiltonian $\op{H}_0$, then the solution $\ket{\Psi}$ of the eigenvalue problem
\begin{align}
\op{H}\ket{\Psi}=E\ket{\Psi}\;,
\end{align}  
can in general not be written in terms of a single \emph{Slater determinant}\index{Slater determinant} $\ket{\occconfig{n}}$ but will be of the form (\ref{superposition}). Besides, those states do not necessarily have a defined particle number. Furthermore, the \emph{Slater determinant} without any particles $\ket{00\ldots}=\ket{\occconfig{0}}$ is the \emph{vacuum state} that is normalized to one according to Eq. (\ref{eq:FockOrthogonal}) and also belongs to the Fock space (The \emph{vacuum state} should not be confused with the zero vector).\par 
Finally, we could have chosen any arbitrary, complete one-particle set $\occconfig{\ket{\nu}}$ in the definition of the states $\ket{\occconfig{n}}$, cf. Eq. (\ref{eq:Besetzungszahlvektor}). Since from the states $\ket{\occconfig{n}}$ it is not clear which one-particle basis has been used for the quantization, one usually has to explicitly specify the chosen basis. 

\subsection{Creation and annihilation operators}
The introduction of creation and annihilation operators, which induce transitions between states of different particle number, is of crucial importance for the formalism of the second quantization. Originally, these operators have been constructed for the description of the harmonic oscillator (so-called ladder operators). In the following bosons and fermions will be treated separately.
\subsubsection{Bosons}
In the bosonic case, the action of the creation and annihilation operators on symmetric states in the occupation number representation is defined by
\begin{align}
\ladderup_i\ket{n_0n_1\ldots n_i\ldots}&=\sqrt{n_i+1}\ket{n_0n_1\ldots n_i+1\ldots}\;,\\
\ladderdown_i\ket{n_0n_1\ldots n_i\ldots}&=\sqrt{n_i}\ket{n_0n_1\ldots n_i-1\ldots}\;,
\end{align}
where the prefactors ensure the correct normalization.  
Apparently, the annihilation operator vanishes for $n_i=0$. Hence, the creation and annihilation operators are not hermitian but pairwise adjoint. The bosonic creation and annihilation operators fulfill the commutation relations
\begin{align}
\kommutator{\ladderup_i,\ladderup_j}&=
\kommutator{\ladderdown_i,\ladderdown_j}=0,\nonumber\\
\kommutator{\ladderdown_i,\ladderup_j}&=\delta_{i,j}\;,
\label{commutation}
\end{align}      
with the commutator $\kommutator{\op{A},\op{B}}=\op{A}\op{B}-\op{B}\op{A}$. Using the creation operator, we can construct arbitrary states of the form (\ref{eq:Besetzungszahlvektor}):
\begin{align}
\ket{\occconfig{n}}=\frac{1}{\sqrt{\prod_i n_i!}}\Bigl(\prod_{i=0}^\infty(\ladderup_i)^{n_i}\Bigr)\ket{\occconfig{0}}.
\end{align}
 Of particular interest is the hermitian occupation number operator of the $i-$th orbital 
 \begin{align}
\op{n}_i=\ladderup_i\ladderdown_i\;. \label{eq:Besetzungszahloperator}
\end{align}
 Its action is given by
 \begin{align}
\op{n}_i\ket{\occconfig{n}}=n_i\ket{\occconfig{n}}\;,
\end{align}
and the eigenvalues are simply the occupation numbers $n_i$ of the $i-$th orbital. Similarly, we have the total particle number operator
\begin{align}
\op{N}=\sum_{i=0}^\infty \op{n}_i\;, \label{eq:Teilchenzahloperator}
\end{align}
with the total particle number $N$ of the state as its eigenvalue
\begin{align}
\op{N}\ket{\occconfig{n}}=N\ket{\occconfig{n}}\;.
\end{align}    
Consequently, the states $\ket{\occconfig{n}}$ are common eigenstates of the occupation number operator, the particle number operator and the ideal Hamilton operator (if we choose the ideal one-particle basis (\ref{idealbases}) for the quantization).  \par
\subsubsection{Fermions}
For fermions, the states have to be antisymmetric. A definition of the fermionic creation and annihilation operators which satisfies this condition is given by
\begin{align}
\ladderup_i \ket{\occconfig{n}}&=(1-n_i)(-1)^{\alpha_{\occconfig{n},i}}\ket{\ldots,n_i+1,\ldots}\;,\nonumber\\  
\ladderdown_i \ket{\occconfig{n}}&=n_i(-1)^{\alpha_{\occconfig{n},i}}\ket{\ldots,n_i-1,\ldots}\;,
\label{def_ladderop}
\end{align}
where the sign is determined by
\begin{align}
\alpha_{\occconfig{n},i} &= \sum_{l=0}^{i-1} n_l\;.
\end{align}
The sign is positive if the number of particles in the one-particle states before the $i-$th state is even (with respect to the chosen order) and negative if it is odd. Further, the prefactors ensure that the creation (annihilation) operator vanishes if the orbital $i$, where the particle shall be created (annihilated), is already occupied (unoccupied), and so the definition also includes the \emph{Pauli principle}. In contrast to the bosonic commutation relations, for fermions, the creation and annihilation operators fulfill the three anti-commutation relations:
\begin{align}
\antikommutator{\ladderup_i,\ladderup_j}&=\antikommutator{\ladderdown_i,\ladderdown_j}=0\;,\nonumber\\
\antikommutator{\ladderdown_i,\ladderup_j}&=\delta_{i,j}\; \label{eq:Antikommutatorrelationen}
\end{align}   
with the anti-commutator being defined as $\antikommutator{\op{A},\op{B}}=\op{A}\op{B}+\op{B}\op{A}$. When it comes to the algebraic treatment of physical systems, the only difference between the second quantization for bosons and for fermions lies in the difference of the commutation relation (\ref{commutation}) and anticommutation relation (\ref{eq:Antikommutatorrelationen}). For example, from the first anti-commutation relation directly follows that $(\op{a}_i^\dagger)^2=0$, i.e. doubly occupied orbitals do not exist in the case of fermions. As a consequence of the non-commutativity of the fermionic creation and annihilation operators, we have to make sure the ordering of the creation operators equals the chosen order of the orbitals when constructing an arbitrary state from the vacuum states according to
\begin{align}
\ket{\occconfig{n}}=\Bigl(\prod_{i=0}^\infty(\ladderup_i)^{n_i}\Bigr)\ket{\occconfig{0}}\;.
\end{align} 
The formulas for the occupation number operator and the particle number operator (\ref{eq:Besetzungszahloperator})-(\ref{eq:Teilchenzahloperator}) remain unchanged for fermions. \par
For further calculations, we will need the matrix elements of the creation and annihilation operators
\begin{align}
\braket{\occconfig{n}|\ladderup_k|\occconfig{\bar{n}}}&=(-1)^{\alpha_{\occconfig{n},k}} \delta_{\occconfig{n},\occconfig{\bar{n}}}^k \delta_{n_k,1}\delta_{n_k,\bar{n}_k+1}\;, \nonumber\\
\braket{\occconfig{n}|\ladderdown_k|\occconfig{\bar{n}}}&=(-1)^{\alpha_{\occconfig{n},k}} \delta_{\occconfig{n},\occconfig{\bar{n}}}^k \delta_{n_k,0}\delta_{n_k,\bar{n}_k-1}\;,\nonumber\\
\delta_{\occconfig{n},\occconfig{\bar{n}}}^k :&= \prod_{\substack{i=0\\i\neq k}}^\infty \delta_{n_i,\bar{n}_i}\;.
\label{matrix_el_ladder}
\end{align}
These follow directly from the definitions (\ref{def_ladderop}).  
Evidently, the creation and annihilation operators are defined with respect to a certain one-particle basis $\occconfig{\ket{i}}$, the basis that we chose for the quantization. The transformation of the creation and annihilation operators to a different basis $\occconfig{\ket{\nu}}$ can be performed as follows:
\begin{align}
\ladderup_\nu&=\sum_{i=0}^\infty\braket{i|\nu}\ladderup_i\;,\\
\ladderdown_\nu&=\sum_{i=0}^\infty\braket{\nu|i}\ladderdown_i\;. \label{eq:TransformationLeiteroperatoren}
\end{align}
For example, choosing spin-coordinate space for the quantization, i.e. $\ket{\nu}=\ket{x}\;,x=\occconfig{\vec{r},\sigma}$, leads to the so-called \emph{field operators}
\begin{align}
\op{\Psi}^\dagger(x):=\op{a}^\dagger_{x}=\sum_{i=0}^\infty\braket{i|x}\op{a}^\dagger_i=\sum_{i=0}^\infty\phi_i^*(x)\op{a}^\dagger_i\,,\\
\op{\Psi}(x):=\op{a}_{x}=\sum_{i=0}^\infty\braket{x|i}\op{a}_i=\sum_{i=0}^\infty\phi_i(x)\op{a}_i\;.
\label{fieldop_def}
\end{align}
The \emph{field operator} $\op{\Psi}^\dagger(x)$ creates and $\op{\Psi}(x)$ annihilates a particle at the space point $\vec{r}$ with spin projection $\sigma$.

\subsection{Operators in second quantization}
Arbitrary first quantized operators can be expressed in the second quantization via the creation and annihilation operators. In the following, we will consider this representation for the one- and two-particle operators. For fermions, in particular, we will give the expressions for the corresponding matrix elements of the interaction potential, which leads to the \emph{Slater-Condon-rules}. 
\subsubsection{One-particle operators}
Though somewhat misleadingly, it is common to refer to the $N-$particle operator 
\begin{align}
\op{B}_1=\sum_{\alpha}\op{b}_\alpha\;,       
\end{align}
that is a sum of true one-particle operators $\op{b}_\alpha$ as a ``one-particle operator'', too. In second quantization, that operator takes the following form:
\begin{align}
\op{B}_1 = \sum_{i,j=1}^\infty b_{ij} \ladderup_i \ladderdown_j, \label{eq:ZweiteQuantisierungEinteilchoperator}
\end{align}
with the one-particle integrals
\begin{align}
b_{ij}=\braket{i|\op{b}|j}=\int \D x \phi_i^*(x) b(x) \phi_j(x)\;,\label{eq:Einteilchenintegrale}
\end{align} 
where the integration over $x=\occconfig{\vec{r},\sigma}$ includes an integration over the space coordinate and a summation over the spin. As usual, we have assumed local operators in coordinate space. From the matrix elements of the creation and annihilation operators (\ref{matrix_el_ladder}), we readily obtain the matrix elements of the one-particle operators (\ref{eq:ZweiteQuantisierungEinteilchoperator}):
\begin{align}
\braket{\occconfig{n}|\ladderup_l\ladderdown_k|\occconfig{\bar{n}}}&= \sum_{\occconfig{n'}} \braket{\occconfig{n}|\ladderup_l|\occconfig{n'}}\braket{\occconfig{n'}|\ladderdown_k|\occconfig{\bar{n}}}\nonumber\\
&=\sum_{\occconfig{n'}}(-1)^{\alpha_{\occconfig{n'},l}+{\alpha_{\occconfig{\bar{n}},k}}} \delta_{\occconfig{n},\occconfig{n'}}^l \delta_{\occconfig{n'},\occconfig{\bar{n}}}^k \delta_{n_l',0} \delta_{n_l,1} \delta_{\bar{n}_k,1}\delta_{n_k',0 }\;.\quad 
\end{align}  
Via case differentiation we can simplify this to
\begin{align}
\braket{\occconfig{n}|\ladderup_l\ladderdown_k|\occconfig{\bar{n}}} =  \begin{cases}
(-1)^{\alpha_{\occconfig{n},l}+{\alpha_{\occconfig{\bar{n}},k}}} \delta_{\occconfig{n},\occconfig{\bar{n}}}^{kl} \delta_{n_l,1}\delta_{\bar{n}_k,1} \delta_{\bar{n}_l,0} \delta_{n_k,0}, \quad & k\neq l \\
n_l\delta_{\occconfig{n},\occconfig{\bar{n}}},\quad & k= l
\end{cases}\;.\label{eq:MatrixelementZweierLeiteroperatoren}
\end{align}
Inserting this into Eq. (\ref{eq:ZweiteQuantisierungEinteilchoperator}) and rearranging the sums with respect to terms with $k=l$ and $k\neq l$ yields
\begin{align}
\braket{\occconfig{n}|\op{B}_1|\occconfig{\bar{n}}} &= \delta_{\occconfig{n},\occconfig{\bar{n}}}\sum_{k=0}^\infty b_{kk} n_k \nonumber\\
&\mathrel{\phantom{=}} + \:\!\sum_{k=1}^\infty \sum_{\substack{l=1\\l\neq k}}^\infty b_{lk}(-1)^{\alpha_{\occconfig{n},l}+{\alpha_{\occconfig{\bar{n}},k}}} \delta_{\occconfig{n},\occconfig{\bar{n}}}^{kl} \delta_{n_l,1}\delta_{\bar{n}_k,1} \delta_{\bar{n}_l,0} \delta_{n_k,0}\;.
\end{align}
The first term only gives a contribution if both \emph{Slater determinants} are equal; thus, these are the diagonal elements of the matrix. The second term does not vanish only if the right and left state differ in exactly two occupation numbers while conserving the total particle number. \par
Let $\ket{\occconfig{n}_q^p}$ be the $N-$particle state that one obtains from the state $\ket{\occconfig{n}}$ if one 
particle is removed from the $q-$th orbital and added to the $p-$th orbital, i.e. for $q<p$ it is 
\begin{align}
\ket{\occconfig{n}_q^p} = \ket{\ldots,n_q-1,\ldots,n_p+1,\ldots}\;. 
\end{align} 
Using this notation, we can rewrite the matrix elements of a one-particle operator of the form (\ref{eq:ZweiteQuantisierungEinteilchoperator}) in a compact way:
\begin{align}
\label{one_particle_op_matrix_elements}
\braket{\occconfig{n}|\op{B}_1|\occconfig{\bar{n}}} = \begin{cases}
\sum\limits_{k=1}^\infty b_{kk}n_k, &\occconfig{n}=\occconfig{\bar{n}} \\[0.2cm]
b_{pq}(-1)^{\sum\limits_{l=\min(p,q)+1}^{\max(p,q)-1}n_l}, &  \occconfig{n}=\occconfig{\bar{n}}_q^p\\[0.2cm]
0, &\text{else}
\end{cases}\;.
\end{align} 
Thereby, we have expressed the matrix elements of a second quantized one-particle operator (\ref{eq:ZweiteQuantisierungEinteilchoperator}) by the one-particle integrals (\ref{eq:Einteilchenintegrale}). 

\subsubsection{Two-particle operators} 
In analogy to the one-particle operators, general two-particle operators of the form $\op{B}_2=\frac{1}{2}\sum_{\alpha\neq\beta=1}^N \op{b}_{\alpha,\beta}$ take the following form in second quantization:
\begin{align}
\op{B}_2=\frac{1}{2}\sum_{i,j,k,l=1}^\infty b_{ijkl} \ladderup_i \ladderup_j \ladderdown_l \ladderdown_k\;.
\end{align}
Note the exchange of the orbital indices of the annihilation operators and the two-particle integrals, which is important for fermions due to the non-commutativity of the creation and annihilation operators. The two-particle integrals are given by
\begin{align}
b_{ijkl}=\braket{ij|\op{b}|kl}= \int \D x \int \D y \,\phi_i^*(x)\phi_j^*(y) b(x,y) \phi_k(x) \phi_l(y)\;.\label{eq:Zweiteilchenintegrale}
\end{align} 
The following considerations are restricted to fermions only. For the most interesting case of the pair interaction operator $\op{W}$, one can take advantage of the fact that the interaction is symmetric under particle exchange, i.e. $w(x,y)=w(y,x)$, and real, i.e. $w^*(x,y)=w(x,y)$. Hence, for the two-particle integrals, we have
\begin{align}
w_{ijkl}&=w_{jilk}\;,\\
w_{ijkl}^*&=w_{klij}\;.
\end{align}
Using these symmetries, we can bring the interaction operator into a more suitable form:
\begin{align}
\W=\sum_{i=1\vphantom{k=1}}^{\infty} \sum_{j=i+1\vphantom{k=1}}^{\infty} \sum_{k=1}^{\infty} \sum_{l=k+1}^{\infty} w_{ijkl}^- \:\ladderup_i \ladderup_j \ladderdown_l \ladderdown_k\;,
\end{align} 
where we have introduced the antisymmetric two-particle integrals $w_{ijkl}^-=w_{ijkl}-w_{ijlk}$. Obviously, there are only terms with $i<j$ and $k<l$. Therefore, we end up with six different cases to be considered:
\begin{align}
\W&= \mathbin{\phantom{+}}\:\! \sum_{i=1} \sum_{j=i+1} w_{ijij}^- \;\ladderup_j \ladderup_i \ladderdown_i \ladderdown_j
+ \sum_{i=1} \sum_{j=i+1} \sum_{\substack{l=i+1\\l\neq j}} w_{ijil}^- \;\ladderup_j \ladderup_i \ladderdown_i \ladderdown_l \nonumber\\
&\mathrel{\phantom{=}} + \:\! \sum_{i=1} \sum_{j=i+1} \sum_{\substack{k=1\\k\neq i}}^{j-1} w_{ijkj}^- \;\ladderup_j \ladderup_i \ladderdown_k \ladderdown_j + \sum_{i=1} \sum_{j=i+1} \sum_{k=1}^{i-1} w_{ijki}^- \;\ladderup_j \ladderup_i \ladderdown_k \ladderdown_i\nonumber \\
&\mathrel{\phantom{=}} + \:\! \sum_{i=1} \sum_{j=i+1} \sum_{l=j+1} w_{ijjl}^- \;\ladderup_j \ladderup_i \ladderdown_j \ladderdown_l +  \sum_{i=1} \sum_{j=i+1} \sum_{\substack{k=1\\k\neq i,j}} \sum_{\substack{l=k+1\\l\neq i,j}} w_{ijkl}^- \;\ladderup_j \ladderup_i \ladderdown_k \ladderdown_l\;.
\end{align}
For each of these cases, one readily computes the matrix elements $\braket{\occconfig{n}|\ladderup_j \ladderup_i \ladderdown_k \ladderdown_l|\occconfig{\bar{n}}}$. Rearranging of the terms finally yields
 \begin{align}
\braket{\occconfig{n}|\W|\occconfig{\bar{n}}}&= \phantom{+} \: \delta_{\occconfig{n},\occconfig{\bar{n}}}\sum_{i=1}^\infty \sum_{j=i+1}^\infty w_{ijij}^- n_i n_j\nonumber \\
&\mathrel{\phantom{=}}+\:\begin{aligned}[t] \sum_{p=1}^\infty \sum_{q=1}^\infty   &\delta_{\occconfig{n},\occconfig{\bar{n}}}^{pq} \delta_{n_p,1} \delta_{\bar{n}_p,0} \delta_{n_q,0} \delta_{\bar{n}_q,1} \nonumber\\
&\cdot\sum_{i=1}^\infty w_{ipiq}^- (-1)^{\alpha_{\occconfig{n},i}+\alpha_{\occconfig{\bar{n}},i}+\alpha_{\occconfig{n},p}+\alpha_{\occconfig{\bar{n}},q}}\Theta(i,p,q)n_i \end{aligned}\nonumber \\
&\mathrel{\phantom{=}}+\: \begin{aligned}[t]\sum_{p=1}^\infty \sum_{q=p+1}^\infty \sum_{r=1}^\infty \sum_{s=r+1}^\infty &\delta_{\occconfig{n},\occconfig{\bar{n}}}^{pqrs} \delta_{n_p,1} \delta_{\bar{n}_p,0} \delta_{n_q,1} \delta_{\bar{n}_q,0} \delta_{n_r,0} \delta_{\bar{n}_r,1} \delta_{n_s,0} \delta_{\bar{n}_s,1}\\
& \cdot w_{pqrs}^- (-1)^{\alpha_{\occconfig{n},p}+\alpha_{\occconfig{n},q}+\alpha_{\occconfig{\bar{n}},r}+\alpha_{\occconfig{\bar{n}},s}}\;, 
\end{aligned}
\end{align}
with
\begin{align}
\Theta(i,p,q)=\begin{cases}
-1, &p<i<q,\, \text{ or } \, q<i<p \\
0, & i=p \,\text{ or }\, i=q\\
1, &\text{else}
\end{cases}\;.
\end{align}
Thus, there are three different contributions to the matrix elements. These are the \emph{Slater-Condon-rules}\index{Slater-Condon rules} \cite{helgaker_molecular_2000}, which can be rewritten as follows:
\begin{align}
\W_{\occconfig{n},\occconfig{\bar{n}}}&\definedby\braket{\occconfig{n}|\W|\occconfig{\bar{n}}} =\W_{\occconfig{n},\occconfig{\bar{n}}}^\mathrm{I}+\W_{\occconfig{n},\occconfig{\bar{n}}}^\mathrm{II}+\W_{\occconfig{n},\occconfig{\bar{n}}}^\mathrm{III}\;,\nonumber\\
\W_{\occconfig{n},\occconfig{\bar{n}}}^\mathrm{I}&=\delta_{\occconfig{n},\occconfig{\bar{n}}}\sum_{i=0}^\infty \sum_{j=i+1}^\infty w_{ijij}^- n_i n_j\;,\nonumber\\
\W_{\occconfig{n},\occconfig{\bar{n}}}^\mathrm{II}&=\sum_{p,q=1}^\infty\delta_{\occconfig{n},\occconfig{\bar{n}}}^{pq}\delta_{n_p,1}\delta_{\bar{n}_p,0}\delta_{n_q,0}\delta_{\bar{n}_q,1}\sum_{\substack{i=0\\i\neq p,q}}w_{ipiq}^- (-1)^{\sum_{l=\min(p,q)+1}^{\max(p,q)-1}n_l}n_i\;,\nonumber\\
\W_{\occconfig{n},\occconfig{\bar{n}}}^\mathrm{III}&=\begin{aligned}[t]\sum_{p=1}^\infty \sum_{q=p+1}^\infty \sum_{r=1}^\infty \sum_{s=r+1}^\infty &\delta_{\occconfig{n},\occconfig{\bar{n}}}^{pqrs} \delta_{n_p,1}\delta_{\bar{n}_p,0}\delta_{n_q,1}\delta_{\bar{n}_q,0} \delta_{n_r,0}\delta_{\bar{n}_r,1}\delta_{n_s,0}\delta_{\bar{n}_s,1} \\
& \cdot w_{pqrs}^- (-1)^{\sum_{l=p}^{q-1}n_l+\sum_{l=r}^{s-1}\bar{n}_l}\;.\label{eq:Slater-Condon}
\end{aligned}
\end{align}
The first term only gives a contribution if both states are equal and the second if left and right state differ in exactly two orbitals. In contrast, the third term is not vanishing only if both states differ in exactly four orbitals. In all cases, the left and right state must have the same particle number. All other matrix elements vanish. Collecting the obtained results, we finally end up with
\begin{align}
\braket{\occconfig{n}|\W|\occconfig{\bar{n}}}=\begin{cases}
\sum_{i=0}^\infty \sum\limits_{j=i+1}^\infty w_{ijij}^- n_i n_j, &\occconfig{n}=\occconfig{\bar{n}}\\[0.2cm]
\sum_{\substack{i=0\\i\neq p,q}}w_{ipiq}^- (-1)^{\sum_{l=\min(p,q)+1}^{\max(p,q)-1}n_l}n_i, &  \occconfig{n}=\occconfig{\bar{n}}_q^p\\[0.2cm]
w_{pqrs}^- (-1)^{\sum_{l=p}^{q-1}n_l+\sum_{l=r}^{s-1}\bar{n}_l}, &  \occconfig{n}=\occconfig{\bar{n}}_{r<s}^{p<q} \\[0.2cm]
0, & \text{else}
\end{cases}\;.
\label{final_slater_rules}
\end{align}
Thereby, $\ket{\occconfig{n}_{r<s}^{p<q}}$ is the $N-$particle state that one obtains from $\ket{\occconfig{n}}$ if a particle is added to the orbitals $\ket{p}$ and $\ket{q}$ each, with $p<q$, and removed from the orbitals $\ket{r}$
and $\ket{s}$, with $r<s$. 

\section{The density operator\label{density_op_chap}}
The density operator $\op{\rho}$ is of extreme importance for the description of thermodynamic properties of quantum many-particle systems. 
While in standard quantum mechanics a quantum system is described by a certain state $\ket{\Psi}$, this does not apply to a quantum system 
at a finite temperature (e.g. a system in a thermostate). In this case, in principle, all possible states may be observed with  a certain 
probability, i.e. the system is in an ensemble of states or in a ``mixed state''.
For an ensemble\index{ensemble} of possible many-particle states $\occconfig{\ket{\Psi_i}}$, the density operator (which replaces the wave function in the previous case of a ``pure'' state) is defined as
\begin{align}
\op{\rho}=\sum_{i}P(\ket{\Psi_i})\ket{\Psi_i}\bra{\Psi_i}\;.
\end{align}
The sum goes over all states of the ensemble, and $P(\ket{\Psi_i})$ is the real, positive probability to observe the state $\ket{\Psi_i}$. The matrix representation of $\op{\rho}$ in an arbitrary N-particle basis $\{\ket{\Phi_i}\}$ is called the density matrix\index{density matrix} $\braket{\Phi|\op{\rho}|\Phi}$. Further, it is hermitian, i.e. $\op{\rho}^\dagger=\op{\rho}$, and positive semi-definite, i.e. $\braket{\Phi_i|\op{\rho}|\Phi_i}\geq0$ for arbitrary states $\ket{\Phi_i}$. The probabilities obey the normalization $ \sum_{i}P(\ket{\Psi_i})=1$, which corresponds to 
\begin{align}
\Tr \op{\rho} = \sum_i \braket{\Phi_i|\op{\rho}|\Phi_i} =1\;,
\end{align}   
where the trace can be evaluated in an arbitrary N-particle basis. The most important feature of the density matrix is that we can compute the expectation value of any observable via   
\begin{align}
\braket{\op{O}}=\Tr (\op{O}\op{\rho})\;.
\label{expectation_value}
\end{align}
It is always $\Tr\op{\rho}^2\leq1$, and for a mixed ensemble this expression is strictly smaller than 1. For a pure state $\ket{\Psi}$, the density operator becomes a projection operator with $\op{\rho}^2=\op{\rho}$ and (\ref{expectation_value}) turns into the standard quantum mechanical average $\braket{\op{O}}=\braket{\Psi|\op{O}|\Psi}$. The entropy of the system can be defined as
\begin{align}
S=-\braket{\ln \op{\rho}}\;.
\end{align}
Hence, the density operator contains the whole thermodynamic information of the system. \par
In the canonical ensemble\index{ensemble!canonical}, characterized by the particle number $N$, the temperature\index{temperature} $T$ and the volume $V$ of the system, the density operator takes the form 
\begin{align}
\op{\rho}(N,\beta,V)=\frac{1}{Z(N,\beta,V)}\:e^{-\beta\Ham}\;,
\end{align}
with the inverse temperature $\beta=\frac{1}{T}$ (we use $k_B=1$). The normalization factor $Z$ is the partition function\index{partition function}
\begin{align}
Z(N,\beta,V)=\Tr e^{-\beta\Ham}\;. \label{eq:KanonischeZustandssumme}
\end{align} 
For most of the physical quantities, the whole information about the $N$-particle density operator is not required. Instead, it is sufficient to know the partition function. Another crucial quantity of the canonical ensemble is the free energy
\begin{align}
F(N,\beta,V)=-T\ln Z\;.
\end{align}  
Most of the physical quantities can also be derived from the free energy, e.g. for the total energy and the heat capacity\index{heat capacity} at constant volume we have
\begin{align}
\braket{\Ham}=-T^2\left.\frac{\partial}{\partial T}\frac{F}{T}\right\vert_{N,V}=-\left.\frac{\partial}{\partial \beta} \ln Z\right\vert_{N,V}\;,\label{eq:AllgemeinEnergie}
\end{align}
and
\begin{align}
C_V=\left.\frac{\partial}{\partial T} \braket{\Ham} \right\vert_{V}\;.\label{eq:AllgemeinCV}
\end{align}
\par
For a system of constant temperature $T$, chemical potential $\mu$ and volume $V$, we use the density operator of the {\em grand canonical ensemble}  
\begin{align}
\op{\rho}(\beta,\mu,V)=\frac{1}{Z(\beta,\mu,V)}e^{-\beta(\Ham-\mu\op{N})},
\end{align}
with the grand canonical partition function 
 \begin{align}
Z(\beta,\mu,V)=\Tr e^{-\beta(\Ham-\mu\op{N})}\;, \label{eq:GrosskanonischeZustandssumme}
\end{align}
 and the grand canonical potential
\begin{align}
\Omega(\beta,\mu,V) = -T\ln Z.
\end{align}
For example, the average particle number and the total energy can be computed according to
\begin{align}
\braket{\op{N}}=-\left.\frac{\partial}{\partial \mu}\Omega\right\vert_{\beta,V}
\;,\qquad
\braket{\Ham}=\left.\frac{\partial}{\partial \beta}\beta\Omega\right\vert_{\beta\mu,V}\;.\label{AllgemeinEnergieGross}
\end{align}
In second quantization, the one-particle density matrix is of particular importance:
\begin{align}
d_{ij}\definedby\braket{\ladderup_i\ladderdown_j}=\Tr \ladderup_i\ladderdown_j \op{\rho}\;.
\end{align}
With the representation of one-particle operators in second quantization (\ref{eq:ZweiteQuantisierungEinteilchoperator}), we can apparently calculate all one-particle quantities from the knowledge of the one-particle density matrix:
\begin{align}
\braket{\op{O}}=\sum_{i,j=0}^\infty o_{ij} \Tr \ladderup_i\ladderdown_j \op{\rho}=\sum_{i,j=0}^\infty o_{ij} d_{ij}\;.
\end{align}
In addition, the diagonal elements of the one-particle density matrix are the average occupation numbers of the orbitals, i.e. $\braket{\op{n}_i}=d_{ii}$. 
For both the canonical and the grand canonical ensembles, the elements of the one-particle density matrix can be obtained from the partition function. In the canonical ensemble it is
\begin{align}
d_{ij}&=\frac{1}{Z}\Tr \ladderup_i\ladderdown_j \:\: e^{-\beta \bigl(\sum\limits_{pq} h_{pq} \ladderup_p\ladderdown_q + \sum\limits_{p<qr<s}w_{pqrs}^-\ladderup_p\ladderup_q\ladderdown_s\ladderdown_r\bigr)}\nonumber\\
&=-\frac{1}{\beta}\frac{\partial}{\partial h_{ij}} \ln Z\;. \label{eq:Einteilchendichtematrix}
\end{align}  
 This relation also holds in the grand canonical ensemble. Moreover, using the definition of the field operators (\ref{fieldop_def}), we can compute the average particle density via
 \begin{align}
n(\vec{r})= \sum_\sigma \braket{\op{\Psi}^\dagger(x)\op{\Psi}(x)}= \sum_\sigma \sum_{i,j=0}^\infty \phi_i(x)^* \phi_j(x) d_{ij}\;.\label{eq:Einteilchendichte}
\end{align}

\section{Configuration path integral Monte Carlo in the interaction representation\label{CPIMC_Chap}}
In this section, first, an expansion of the partition function suitable for a CPIMC algorithm shall be derived. Next, we will show how to obtain the estimators of physical quantities and explain the used Monte Carlo steps in detail. Finally, some results will be shown for a one-dimensional system of Coulomb-interacting electrons in a harmonic trap. 
 
\subsection{Expansion of the partition function}
From now on, we will only consider fermions in the canonical ensemble. Hence, the partition function is given by (\ref{eq:KanonischeZustandssumme}).
There are many different ways to expand this expression. First, we have to choose a certain basis in which we evaluate the trace, and second, there are quite a few possibilities to factorize the density operator. Each combination of basis and factorization will lead to a different, more or less efficient\footnote{Efficient with respect to a certain system with fixed parameters to be simulated, i.e. currently, there is no method suitable for all systems. }, Monte Carlo algorithm. For example, we could choose coordinate space and use the \emph{Trotter formula} \cite{trotter_product_1959}, \cite{suzuki_generalized_1976}:
\begin{align}
e^{\beta\op{H}}= \lim_{n\to \infty} \left(e^{\frac{\beta}{n}\op{H}_0}e^{\frac{\beta}{n}\op{W}}\right)^n\;.
\end{align}
In that case, we have to antisymmetrize the density operator which yields the DPIMC method \cite{ceperley_path_1995}. We could also perform the trace in occupation number representation, which will automatically ensure the correct symmetrization. In combination with the standard expansion of the exponential function
\begin{align}
e^{x}=\sum_{k=0}^\infty \frac{x^k}{k!}\;,
\end{align}  
we would end up with the method called \emph{Stochastic Series Expansion} \cite{Sandvik_SSE_1991}. \par
Here, we follow a different approach based on the ideas of N.V. Prokof'ev {\em et al.} \cite{prokofev_exact_1998} and\cite{prokofev_trieste_2008}. First, we split the Hamiltonian $\op{H}$ into a diagonal part $\op{D}$ and an off-diagonal part $\op{Y}$, i.e. for an arbitrary basis $\occconfig{\ket{\alpha}}$ it is
\begin{align}
\braket{\alpha|\op{H}|\alpha^\prime}=\begin{cases}
\braket{\alpha|\op{D}|\alpha}=D_{\alpha,\alpha}\,,\qquad&\text{if}\qquad \alpha=\alpha^\prime \\
\braket{\alpha|\op{Y}|\alpha^\prime}=Y_{\alpha,\alpha^\prime}\,,\qquad&\text{if}\qquad \alpha\neq\alpha^\prime\end{cases}\;.
\label{splitting}
\end{align} 
Second, we switch to the interaction picture in imaginary time with respect to $\op{D}$ and make use of the well-known identity\footnote{In this notation, operators in the interaction and the Schr\"odinger picture differ only concerning the presence of a time argument.}
\begin{align}
  &e^{-\beta\op{H}}\equiv e^{-\beta\op{D}}\op{T}_\tau \: e^{-\int_0^\beta\op{Y}(\tau)\mathrm{d}\tau}\:, \quad\text{with}\quad \op{Y}(\tau)=e^{\tau\op{D}}\op{Y}e^{-\tau\op{D}},\;\; \tau\in(0,\beta)\;,
 \label{identity}
\end{align}
where the action of the time-ordering operator $\op{T}_\tau$ on two arbitrary operators $\op{A}(\tau_1)$ and $\op{B}(\tau_2)$ is defined by
\begin{align}
\op{T}_\tau[\op{A}(\tau_1)\op{B}(\tau_2)]=\begin{cases}
\op{A}(\tau_1)\op{B}(\tau_2)\;,\qquad&\text{if}\qquad \tau_1>\tau_2\\
\op{A}(\tau_2)\op{B}(\tau_1)\;,\qquad&\text{if}\qquad \tau_2>\tau_1
\end{cases}\;.
\end{align} 
Next, we expand the exponential function of the off-diagonal operator:
\begin{align}
  \op{T}_\tau e^{-\int_0^\beta\op{Y}(\tau)\mathrm{d}\tau}&=\sum_{K=0}^\infty\int_0^\beta \mathrm{d}\tau_1\ldots \int_0^\beta \mathrm{d}\tau_K\frac{(-1)^K}{K!}\op{T}_\tau[\op{Y}(\tau_1)\op{Y}(\tau_2)\cdot\ldots\cdot\op{Y}(\tau_K)]\\
 &=\sum_{K=0}^\infty\tauint(-1)^K\op{Y}(\tau_K)\op{Y}(\tau_{K-1})\cdot\ldots\cdot\op{Y}(\tau_1)\nonumber\;.
 \label{exponential_expansion}
\end{align}
In the first line, each integral goes from $0$ to $\beta$. Since the time-ordering operator always arranges those off-diagonal operators with the latest times to the left, for each $K$, there are $K!$ equal terms. Therefore, we can eliminate the time-ordering operator by canceling the $K!$, modifying the integral borders and arranging the operator product in such a way that it is already time-ordered.\par
Combining Eq. (\ref{eq:KanonischeZustandssumme}), (\ref{identity}) and (\ref{exponential_expansion}) yields
\begin{align}
  Z&=\sum_{K=0}^\infty\sum_{\occconfig{n}}\tauint\nonumber\\
  &\hspace{3cm}(-1)^Ke^{-\beta\op{D}_{\occconfig{n}}}\braket{\occconfig{n}|\op{Y}(\tau_K)\op{Y}(\tau_{K-1})\cdot\ldots\cdot\op{Y}(\tau_1)|\occconfig{n}}\;,
  \end{align} 
where we chose the occupation number representation to perform the trace, i.e. in Eq. (\ref{splitting}) it is $\ket{\alpha}=\ket{\occconfig{n}}$. Now we insert $K-1 $ unit operators of the form $\op{1}=\sum_{\occconfig{n^{(i)}}}\ket{\occconfig{n^{(i)}}}\bra{\occconfig{n^{(i)}}}$ in between the off-diagonal operators. By definition of the interaction picture, cf. Eq. (\ref{identity}), the resulting matrix elements are of the form
 \begin{align}
\braket{\occconfig{n^{(i)}}|\op{Y}(\tau_K)|\occconfig{n^{(j)}}}&=\braket{\occconfig{n^{(i)}}|e^{\tau_K \op{D}}\op{Y}e^{-\tau_K\op{D}}|\occconfig{n^{(j)}}}\nonumber\\
&=e^{\tau_K D_{\occconfig{n^{(i)}}}}Y_{\occconfig{n^{(i)}},\occconfig{n^{(j)}}}e^{-\tau_K D_{\occconfig{n^{(j)}}}}\label{element}\;.
\end{align}
After rearranging all factors and taking into account that $Y_{\occconfig{n^{(i)}},\occconfig{n^{(i)}}}\equiv 0$, we finally end up with 
\begin{align}
Z(N,V,\beta)=\sum_{K=0,\atop K \neq 1}^{\infty} \sum_{\{n\},\atop =\{n^{(0)}\}=\{n^{(K)}\}} 
\sum_{\{n^{(1)}\},\atop \neq\{n^{(0)}\}} \ldots \sum_{\{n^{(K-1)}\},\atop \neq\{n^{(K-2)}\}\neq\{n^{(K)}\}}
\int\limits_{0}^{\beta} d\tau_1 \int\limits_{\tau_1}^{\beta} d\tau_2 \ldots \int\limits_{\tau_{K-1}}^\beta d\tau_K\nonumber\\
(-1)^K \exp{\left\{-\sum_{i=0}^{K} D_{\{n^{(i)}\}} (\tau_{i+1}-\tau_i)\right\}} \prod_{i=0}^{K-1}
Y_{\{n^{(i)}\},\{n^{(i+1)}\}}=:\intsum_{C}W(C)\label{CZ}\;,
\end{align} 
where the boundary conditions $\occconfig{n}=\occconfig{n^{(0)}}=\occconfig{n^{(K)}}$, $\tau_0=0$ and $\tau_{k+1}=\beta$ hold. For the CPIMC Monte Carlo algorithm, this representation of the partition function\index{partition function} is used. Hence, we can identify the introduced weights $W$ and configurations $C$ of Sec. \ref{Monte_Carlo_chap} in the following way. A contribution to the partition function is uniquely defined by the number of kinks $K$, $K$ kink times $\tau_i$ and $K$ \emph{Slater determinants} which are realized in between the kinks:
\begin{align}
C=\{K, \tau_1,\tau_2, \ldots, \tau_K,
 \occconfig{n^{(0)}},\occconfig{n^{(1)}}, \ldots,\occconfig{n^{(K-1)}}\}\;.
\end{align}
Thus, the weights defined by Eq. (\ref{CZ}) are given by
\begin{align}
\label{weight}
W\Big(C&=\{K, \tau_1,\tau_2, \ldots, \tau_K,
 \occconfig{n^{(0)}},\occconfig{n^{(1)}}, \ldots,\occconfig{n^{(K-1)}}\}\Big)=\\
&\hspace{2cm}(-1)^K \exp{\left\{-\sum_{i=0}^{K} D_{\{n^{(i)}\}} (\tau_{i+1}-\tau_i)\right\}} \prod_{i=0}^{K-1}
Y_{\{n^{(i)}\},\{n^{(i+1)}\}}.\nonumber 
\end{align}
\begin{figure}[t]
\label{Possible_path}
\center{\begin{tikzpicture}[xscale=0.7, yscale=0.7]
\newcommand{\xrange}{10}
\newcommand{\yrange}{4}
\newcommand{\M}{20}
\newcommand{\taueins}{\xrange*0.2}
\newcommand{\tauzwei}{\xrange*0.25}
\newcommand{\taudrei}{\xrange*0.4}
\newcommand{\tauvier}{\xrange*0.7}
\newcommand{\taufuenf}{\xrange*0.9}
\newcommand{\nnull}{\yrange*0.6}
\newcommand{\neins}{\yrange*0.1}
\newcommand{\nzwei}{\yrange*0.4}
\newcommand{\ndrei}{\yrange*0.9}
\newcommand{\nvier}{\yrange*0.4}
\draw[->] (0,0) -- ++(\xrange+0.5*\xrange/\M,0) coordinate (xlabel) ;
\draw[->] (0,0) -- +(0,\yrange) coordinate (ylabel)  ;

\draw[semithick,dotted] (\taueins,-0.1) -- (\taueins,\neins);
\draw[semithick,dotted] (\tauzwei,-0.1) -- (\tauzwei,\neins);
\draw[semithick,dotted] (\taudrei,-0.1) -- (\taudrei,\nzwei);
\draw[semithick,dotted] (\tauvier,-0.1) -- (\tauvier,\nzwei);
\draw[semithick,dotted] (\taufuenf,-0.1) -- (\taufuenf,\nnull);
\draw[semithick,dotted] (\xrange,-0.1) -- (\xrange,\nnull);
\node[rotate=90] at (-3.5,0.5*\yrange) {state $\occconfig{n}(\tau)$};
\node [anchor=east] at (0,\nnull) {$\occconfig{n^{(0)}}$};
\node [anchor=east] at (0,\neins) {$\occconfig{n^{(1)}}$};
\node [anchor=east] at (0,\nzwei) {$\occconfig{n^{(2)}},\occconfig{n^{(4)}}$};
\node [anchor=east] at (0,\ndrei) {$\occconfig{n^{(3)}}$};
\node [anchor=east] at (0,\nvier) {$\occconfig{n^{(4)}}$};
\draw[semithick,dotted] (-0.1,\neins) -- (\taueins,\neins);
\draw[semithick,dotted] (-0.1,\nnull) -- (\taufuenf,\nnull);
\draw[semithick,dotted] (-0.1,\nzwei) -- (\tauvier,\nzwei);
\draw[semithick,dotted] (-0.1,\ndrei) -- (\taudrei,\ndrei);
\node at (0,-0.5) {$0$};
\node at (\taueins,-0.5) {$\tau_1$};
\node at (\tauzwei,-0.5) {$\tau_2$};
\node at (\taudrei,-0.5) {$\tau_3$};
\node at (\tauvier,-0.5) {$\tau_4$};
\node at (\taufuenf,-0.5) {$\tau_5$};
\node at (\xrange,-0.5) {$\beta$};
\node at (0.5*\xrange,-1.2) {imaginary time $\tau$};
\draw[thick,black] (0,\nnull) -- (\taueins,\nnull);
\draw[thick] (\taueins,\nnull) -- (\taueins,\neins) -- (\tauzwei,\neins) -- (\tauzwei,\nzwei) -- (\taudrei,\nzwei) --(\taudrei,\ndrei)--(\tauvier,\ndrei);
\draw[thick] (\tauvier,\ndrei) -- (\tauvier,\nvier);
\draw[thick, black] (\tauvier,\ndrei) -- (\tauvier,\nvier);
\draw[thick,] (\tauvier,\nvier)-- (\taufuenf, \nvier) -- (\taufuenf,\nnull) -- (\xrange,\nnull) ;

\node at (1,0.73*\yrange) {$\textcolor{black}{D_{\occconfig{n}^{(0)}}}$};
\node at (7,4.2) {$\textcolor{black}{Y_{\occconfig{n}^{(3)}\occconfig{n}^{(4)}}}$};
\end{tikzpicture}}
\caption{Possible path in imaginary time. Horizontal lines correspond to diagonal matrix elements, whereas vertical lines correspond to off-diagonal elements. }
\end{figure}
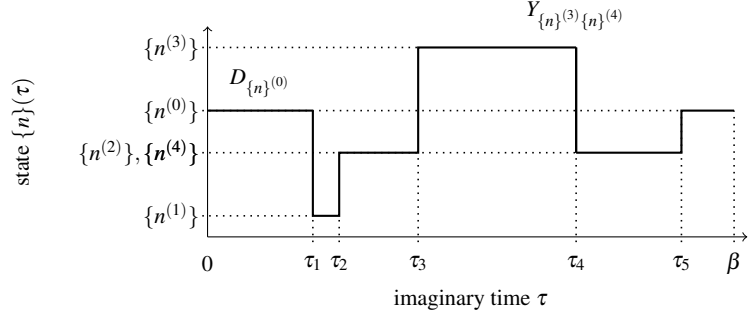
Such configurations can also be visualized as $\beta-$periodic paths in imaginary time.
As an example, Fig. \ref{Possible_path} shows a possible path. Each horizontal line corresponds to a diagonal matrix element in the weight of that configuration [cf. Eq. (\ref{weight})], whereas vertical lines correspond to off-diagonal elements.\par In addition to the correct statistics included in the occupation number representation, this expansion of the partition function yields paths that are continuous in imaginary time. This is different to DPIMC, where the corresponding paths are discrete and one has to ensure the convergence of the discretization by performing several simulations for the same system parameters but different discretizations (different numbers of ``time slices''). For this reason, the presented CPIMC method belongs to the so-called continuous time QMC methods. A different, independent derivation of (\ref{CZ}), that is based on the \emph{Trotter Formula}, was given in \cite{schoof_configuration_2011} where the notion configuration PIMC was introduced.
\par Further, since we have only expanded the exponential function of the off-diagonal 
operator [cf. Eq. (\ref{exponential_expansion})], this representation of the partition function is nothing but a perturbation expansion with respect to the off-diagonal matrix elements. Hence, if we choose the ideal basis for the quantization, it is indeed a perturbation expansion with respect to the interaction.\par
Combining the \emph{Slater-Condon-rules} (\ref{final_slater_rules}) and Eq. (\ref{one_particle_op_matrix_elements}), the matrix elements of the diagonal and off-diagonal operator are calculated according to\footnote{Due to the finite number of one-particle states in a practical simulation, the sums will not go up to infinity but to $N_B$, the number of orbitals used in the simulation. } 
\begin{align}
D_{\occconfig{n^{(k)}}}&=\sum_{i=1}^\infty b_{ii}n^{(k)}_i+\sum_{i=0}^\infty \sum_{j=i+1}^\infty w_{ijij}^- n^{(k)}_i n^{(k)}_j\,,
\end{align}
and
\begin{align}
Y_{\occconfig{n^{(k)}}\occconfig{n^{(l)}}}=\begin{cases}
\displaystyle \Biggl(b_{pq}+\sum_{\substack{i=0\\i\neq p,q}}^\infty w_{ipiq}^-n^{(k)}_i\Biggr)(-1)^{\sum_{m=\min(p,q)+1}^{\max(p,q)-1}n^{(k)}_m} ,&  \occconfig{n^{(k)}}=\occconfig{n^{(l)}}_q^p\\[5ex]
w_{pqrs}^- (-1)^{\sum_{m=p}^{q-1}n^{(k)}_m+\sum_{m=r}^{s-1}n^{(l)}_m},&  \occconfig{n^{(k)}}=\occconfig{n^{(l)}}_{r<s}^{p<q} \\[0.2cm]
0, & \text{else.}
\end{cases}
\label{off_matrix_elments}
\end{align} 
Obviously, only those configurations (paths) contribute to the partition function (\ref{CZ}) in which consecutive \emph{Slater determinants} differ in either two or  one orbitals, since otherwise the off-diagonal matrix elements vanish. Moreover, there are three sources of sign changes in the weights meaning that the fermion \emph{sign problem} (see Sec. \ref{sign_prob_chap}) survives. First, an increment of the number of kinks changes the sign due to the factor $(-1)^K$. Second, we have a phase factor in the off-diagonal matrix elements (\ref{off_matrix_elments}) and third, the one and two particle integrals can have positive and negative signs.

\subsection{Estimators for many-particle observables}
The simplest way to obtain the estimators (cf. Sec. \ref{Monte_Carlo_chap}) is to use the thermodynamic relations of Sec. \ref{density_op_chap} between the partition function and the observables. Sometimes those \emph{trivial estimators} can be numerically unstable, and one has to use different estimators, e.g. in DPIMC  the \emph{virial estimator} for the energy \cite{bonitz_introduction_2006}. In CPIMC, the \emph{trivial estimators} are numerically stable.\par
Starting with the energy, we have to compute $\braket{\op{H}}=-\frac{1}{Z}\frac{\partial Z}{\partial \beta}$ with the partition function (\ref{CZ}). Since both the integrand and the integral borders in (\ref{CZ}) depend on the inverse temperature $\beta$, we substitute $t_i=\frac{\tau_i}{\beta}$ and $d\tau_i=\beta d t_i$ yielding
\begin{align}
Z(N,V,\beta)=\sum_{K=0,\atop K \neq 1}^{\infty} \sum_{\{n\},\atop =\{n^{(0)}\}=\{n^{(K)}\}} 
\sum_{\{n^{(1)}\},\atop \neq\{n^{(0)}\}} \ldots \sum_{\{n^{(K-1)}\},\atop \neq\{n^{(K-2)}\}\neq\{n^{(K)}\}}
\int\limits_{0}^{1} dt_1 \int\limits_{t_1}^{1} dt_2 \ldots \int\limits_{t_{K-1}}^1 dt_K\nonumber\\
(-\beta)^K \exp{\left\{-\beta\sum_{i=0}^{K} D_{\{n^{(i)}\}} (t_{i+1}-t_i)\right\}} \prod_{i=0}^{K-1}
Y_{\{n^{(i)}\},\{n^{(i+1)}\}}\;.
\end{align}           
Now the integral limits are independent of $\beta$ ($t_i\in[0,1]$) and we can easily perform the differentiation:
\begin{align}
\braket{\op{H}}&=\frac{1}{Z}\sum_{K=0,\atop K \neq 1}^{\infty} \sum_{\{n\},\atop =\{n^{(0)}\}=\{n^{(K)}\}} 
\sum_{\{n^{(1)}\},\atop \neq\{n^{(0)}\}} \ldots \sum_{\{n^{(K-1)}\},\atop \neq\{n^{(K-2)}\}\neq\{n^{(K)}\}}\nonumber
\int\limits_{0}^{1} dt_1 \int\limits_{t_1}^{1} dt_2 \ldots \int\limits_{t_{K-1}}^1 dt_K\\\nonumber
&\qquad\qquad\left[-K\beta^{K-1}+\beta^K\sum_{i=0}^{K} D_{\{n^{(i)}\}} (t_{i+1}-t_i)\right]\\\ &\qquad\qquad\times(-1)^K \exp{\left\{-\beta\sum_{i=0}^{K} D_{\{n^{(i)}\}} (t_{i+1}-t_i)\right\}} \prod_{i=0}^{K-1}
Y_{\{n^{(i)}\},\{n^{(i+1)}\}}\;,
\end{align}
and, after substituting back, we obtain the estimator for the total (internal) energy
\begin{align}
\braket{\op{H}}=\frac{1}{Z}\sum_{K=0,\atop K \neq 1}^{\infty} \sum_{\{n\},\atop =\{n^{(0)}\}=\{n^{(K)}\}} 
\sum_{\{n^{(1)}\},\atop \neq\{n^{(0)}\}} \ldots \sum_{\{n^{(K-1)}\},\atop \neq\{n^{(K-2)}\}\neq\{n^{(K)}\}}
\int\limits_{0}^{\beta} d\tau_1 \int\limits_{\tau_1}^{\beta} d\tau_2 \ldots \int\limits_{\tau_{K-1}}^\beta d\tau_K\nonumber\\
\left[\frac{-K}{\beta}+\sum_{i=0}^{K} D_{\{n^{(i)}\}} \frac{(\tau_{i+1}-\tau_i)}{\beta}\right]W(C).
\end{align}  
Thus, the energy of a configuration $C=\{K, \tau_1,\tau_2, \ldots, \tau_K,
 \occconfig{n^{(0)}},\occconfig{n^{(1)}}, \ldots,\occconfig{n^{(K-1)}}\}$ is given by
 \begin{align}
E(C)=\left[\frac{-K}{\beta}+\sum_{i=0}^{K} D_{\{n^{(i)}\}} \frac{(\tau_{i+1}-\tau_i)}{\beta}\right]. 
 \end{align}
 It is worth mentioning that the contribution of the off-diagonal elements of the Hamiltonian to the total energy is given only by the number of kinks multiplied by the temperature. On the other hand, the form of the diagonal contribution is reasonable as well. It is the sum of all diagonal elements of all realized \emph{Slater determinants} weighted with the relative length. \par
According to Eq. (\ref{eq:AllgemeinCV}), the estimator for the heat capacity\index{heat capacity} takes the following form:
\begin{align}
C_V&=\begin{aligned}[t]\frac{1}{T^2}\frac{1}{Z}&\sum_{\substack{K=0\\K\neq 1}}^\infty\;  \sum_{\substack{\occconfig{n}\\=\occconfig{n^{(0)}}=\occconfig{n^{(K)}}}} \; \sum_{\substack{\occconfig{n^{(1)}}\\\neq\occconfig{n^{(0)}}}} \quad \ldots \quad\smashoperator{\sum_{\substack{\occconfig{n^{(K-1)}}\\\neq\occconfig{n^{(K-2)}},\neq\occconfig{n^{(K)}}}}}^{\hphantom{\neq\occconfig{n^{(0)}}}}\int_0^\beta \mathrm{d}\tau_1 \int_{\tau_1}^\beta \mathrm{d}\tau_2 \ldots \int_{\tau_{K-1}}^\beta \!\! \mathrm{d} \tau_K\\
&\Bigg[\biggl(\frac{1}{\beta} \sum_{i=0}^K D_{\occconfig{n^{(i)}}}(\tau_{i+1}-\tau_i) -\frac{K}{\beta}-\braket{H}\biggr)^2 -\frac{K}{\beta^2}\Bigg]W(C)\;.
\label{eq:KinkKontinuierlicheWärmekapazität}
\end{aligned}
\end{align} 
Other important quantities are the one-particle density matrix and, in particular, its diagonal elements--the average occupation numbers. Using Eq. (\ref{eq:Einteilchendichtematrix}), one readily computes the corresponding estimators:
\begin{align}
d_{pq}&=\frac{1}{Z}\sum_{\substack{K=0\\K\neq 1}}^\infty\;  \sum_{\substack{\occconfig{n}\\=\occconfig{n^{(0)}}=\occconfig{n^{(K)}}}} \; \sum_{\substack{\occconfig{n^{(1)}}\\\neq\occconfig{n^{(0)}}}} \quad \ldots \quad\smashoperator{\sum_{\substack{\occconfig{n^{(K-1)}}\\\neq\occconfig{n^{(K-2)}},\neq\occconfig{n^{(K)}}}}}^{\hphantom{\neq\occconfig{n^{(0)}}}}\qquad \; \int_0^\beta \mathrm{d}\tau_1 \int_{\tau_1}^\beta \mathrm{d}\tau_2 \ldots \int_{\tau_{K-1}}^\beta \!\! \mathrm{d} \tau_K\nonumber\\[0.5\jot]
&\hphantom{=\frac{1}{Z}}\biggl(-\frac{1}{\beta} \sum_{i=0}^{K-1} \frac{\braket{\occconfig{n^{(i)}}|\ladderup_p\ladderdown_q|\occconfig{n^{(i+1)}}}}{Y_{\occconfig{n^{(i)}},\occconfig{n^{(i+1)}}}}  \biggr)W(C)
+\braket{\op{n}_p}\delta_{pq}\;, \label{eq:KinkEinteilchendichtematrix}
\end{align} 

\begin{align}
\braket{\op{n}_p}&=\frac{1}{Z}\sum_{\substack{K=0\\K\neq 1}}^\infty\;  \sum_{\substack{\occconfig{n}\\=\occconfig{n^{(0)}}=\occconfig{n^{(K)}}}} \; \sum_{\substack{\occconfig{n^{(1)}}\\\neq\occconfig{n^{(0)}}}} \quad \ldots \quad\smashoperator{\sum_{\substack{\occconfig{n^{(K-1)}}\\\neq\occconfig{n^{(K-2)}},\neq\occconfig{n^{(K)}}}}}^{\hphantom{\neq\occconfig{n^{(0)}}}}\qquad \; \int_0^\beta \mathrm{d}\tau_1 \int_{\tau_1}^\beta \mathrm{d}\tau_2 \ldots \int_{\tau_{K-1}}^\beta \!\! \mathrm{d} \tau_K\nonumber\\[0.5\jot]
&\hphantom{=\frac{1}{Z}}\bigg( \frac{1}{\beta} \sum_{i=0}^Kn_{p}^{(i)}(\tau_{i+1}-\tau_i)\bigg)W(C)\;.
\end{align}
The average occupation of the $p-$th orbital in the configuration $C$
\begin{align}
n_p(C)=\sum_{i=0}^{K} n_p^{(i)} \frac{(\tau_{i+1}-\tau_i)}{\beta}\,,
\end{align}  
 is nothing but the average of the occupation of the $p-$th orbital over realized \emph{Slater determinants}, again weighted with the relative length of the path  on which the determinant is realized. 

\subsection{CPIMC procedure: Monte Carlo steps\label{steps}}
When developing a Monte Carlo algorithm, the most difficult task lies in the construction of a good set of Monte Carlo steps that is most efficient while simultaneously ensuring the ergodicity. In the program, we start from an initial configuration with zero kinks. Thus, the path consists of only one \emph{Slater determinant} realized from $0$ to $\beta$, i.e. it is $C_0=\{0,\occconfig{n^{0}}\}$. If we simulate $N$ particles, then we populate the first $N$ one-particle orbitals of $\ket{\occconfig{n^{(0)}}}$ while the remaining $(N_B-N)$ orbitals are initially unoccupied. Now we need a sufficient number of Monte Carlo steps that allow for sampling of the whole configuration space\index{configuration space} (all possible paths) in an acceptable amount of computation time. As a consequence of the summation constraint in the partition function~(\ref{CZ}), configurations with only a single kink are forbidden. Hence, to reach configurations with more than zero kinks, we will not only need a step in which a single kink is added but also a step in which two kinks are added at once. \par
Obviously, we also need a step that changes the occupation of the realized \emph{Slater determinants}. One might expect that an implementation of a step that swaps the occupation of an occupied and an unoccupied orbital of a determinant (one-particle excitation) is sufficient. Actually, it can be shown that for a fixed number of kinks the configuration space separates into subspaces, and hence, ergodicity is not fulfilled \cite{Tim_dip}. Meaning, if we only use the one-particle excitation, we will be stuck in a certain subspace, which results in wrong expectation values. The number of separated subspaces for a fixed number of kinks can be reduced by the implementation of a two particle excitation. \par
Since the off-diagonal elements vanish if the two determinants differ in more than four orbitals (cf. Eq.~(\ref{off_matrix_elments})), a step in which three particles are excited will in most cases be rejected. \par
Since we cannot exclude that a separation of the configuration space occurs for a variable number of kinks (thus destroying ergodicity of our procedure), we assign a so-called \emph{virtual weight} to certain forbidden paths. This is achieved by setting each off-diagonal matrix element $Y_{\occconfig{n}\occconfig{\bar{n}}}<Y_{\text{min}}=10^{-10}$, with $\occconfig{n}=\occconfig{\bar{n}}^{p<q}_{r<s}$ to a fixed \emph{virtual weight} $Y_V$. Thereby, the subspaces of the configuration space become connected via the forbidden configurations. If all forbidden paths are neglected in the calculation of the expectation values, then these will still be correct. During the equilibration time of the \emph{Markov process}, the virtual weight will be determined in such a way that between $70$ to $90\%$ off all configurations in the \emph{Markov chain} are allowed. As a consequence, the runtime of the program increases by up to $30\%$ by the use of \emph{virtual weights}. On the other hand, the autocorrelation time of those configurations contributing to the measurements is reduced, and so the extended runtime is partially compensated.\par
A particular step to remove a kink is not needed. Whenever a determinant, after a one- or two-particle excitation has been applied to it, equals its left or right neighboring determinant (in imaginary time), the corresponding kink will be removed. Besides, a particular step to remove kinks would have a factor in its acceptance probability that decreases with the basis size $N_B$. 
\par
Finally, we need a step to shift the position of a kink (in imaginary time). In fact, this step might not be necessary for the ergodicity, since we can remove a certain kink and add it at a different time. However, a particular step in which a kink is shifted helps reducing the autocorrelation time.
\par 
In the following, the Monte Carlo steps shall be explained in detail following \cite{Tim_dip} and \cite{schoof_configuration_2011}. As an illustration, Fig. {\ref{fig:MonteCarloSchritte}} shows the changes in the path for each step. Nevertheless, the $\beta-$periodicity has not been taken into account explicitly. Instead there are two intervals $[0,\tau_1]$ and $[\tau_K, \beta]$ which are both assigned to the same \emph{Slater determinant}:       
     
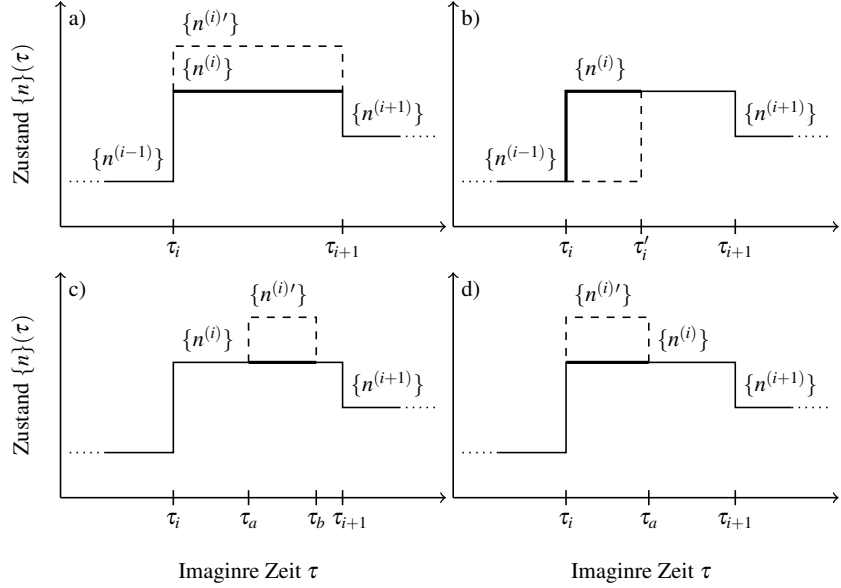
\begin{figure}[t]
\centering
  \mbox{\begin{tikzpicture}[semithick]
\newcommand{\xrange}{5}
\newcommand{\yrange}{3}
\newcommand{\M}{20}
\newcommand{\tauzwei}{\xrange*0.12}
\newcommand{\taudrei}{\xrange*0.3}
\newcommand{\tauvier}{\xrange*0.75}
\newcommand{\taufuenf}{\xrange*0.9}
\newcommand{\nzwei}{\yrange*0.2}
\newcommand{\ndrei}{\yrange*0.6}
\newcommand{\nvier}{\yrange*0.4}
\newcommand{\nnew}{\yrange*0.8}
\draw[->] (0,0) -- ++(\xrange+0.5*\xrange/\M,0) coordinate (xlabel) ;
\draw[->] (0,0) -- +(0,\yrange) coordinate (ylabel)  ;
\draw (\tauzwei,\nzwei) -- (\taudrei,\nzwei) --(\taudrei,\ndrei) --(\tauvier,\ndrei) -- (\tauvier,\nvier) -- (\taufuenf, \nvier) ;
\draw[very thick] (\taudrei,\ndrei) --(\tauvier,\ndrei);
\draw[dotted] (\tauzwei,\nzwei) -- +(-0.5,0);
\draw[dotted] (\taufuenf,\nvier) -- + (0.5,0);
\draw[dashed] (\taudrei,\ndrei) -- (\taudrei,\nnew) -- (\tauvier,\nnew) -- (\tauvier,\ndrei);
\node[below] at (\taudrei,0) {\strut$\tau_i$};
\node[below] at (\tauvier,0) {\strut$\tau_{i+1}$};
\draw (\taudrei,-0.1) -- (\taudrei,0.1);
\draw (\tauvier,-0.1) -- (\tauvier,0.1);
\node[above right] at (\taudrei,\ndrei) {\strut$\occconfig{n^{(i)}}$};
\node[above right] at (\taudrei,\nnew) {\strut$\occconfig{{n^{(i)}}\vphantom{n}'}$};
\node[above left] at (\taudrei,\nzwei) {\strut$\occconfig{{n^{(i-1)}}}$};
\node[above right] at (\tauvier,\nvier) {\strut$\occconfig{n^{(i+1)}}$};
\node[rotate=90] at (-0.5,0.5*\yrange) {Zustand $\occconfig{n}(\tau)$};
\node [below right] at (0,\yrange) {a)};
\end{tikzpicture}\begin{tikzpicture}[semithick]
\newcommand{\xrange}{5}
\newcommand{\yrange}{3}
\newcommand{\M}{20}
\newcommand{\tauzwei}{\xrange*0.12}
\newcommand{\taudrei}{\xrange*0.3}
\newcommand{\tauvier}{\xrange*0.75}
\newcommand{\taufuenf}{\xrange*0.9}
\newcommand{\taua}{\xrange*0.5}
\newcommand{\nzwei}{\yrange*0.2}
\newcommand{\ndrei}{\yrange*0.6}
\newcommand{\nvier}{\yrange*0.4}
\draw[->] (0,0) -- ++(\xrange+0.5*\xrange/\M,0) coordinate (xlabel) ;
\draw[->] (0,0) -- +(0,\yrange) coordinate (ylabel)  ;
\draw (\tauzwei,\nzwei) -- (\taudrei,\nzwei) --(\taudrei,\ndrei) --(\tauvier,\ndrei) -- (\tauvier,\nvier) -- (\taufuenf, \nvier) ;
\draw[very thick] (\taudrei,\nzwei) --(\taudrei,\ndrei) --(\taua,\ndrei);
\draw[dotted] (\tauzwei,\nzwei) -- +(-0.5,0);
\draw[dotted] (\taufuenf,\nvier) -- + (0.5,0);
\draw[dashed] (\taudrei,\nzwei) -- (\taua,\nzwei) -- (\taua,\ndrei);
\node[below] at (\taudrei,0) {\strut$\tau_i$};
\node[below] at (\tauvier,0) {\strut$\tau_{i+1}$};
\node[below] at (\taua,0) {\strut$\tau_{i}'$};
\draw (\taudrei,-0.1) -- (\taudrei,0.1);
\draw (\tauvier,-0.1) -- (\tauvier,0.1);
\draw (\taua,-0.1) -- (\taua,0.1);
\node[above right] at (\taudrei,\ndrei) {\strut$\occconfig{n^{(i)}}$};
\node[above left] at (\taudrei,\nzwei) {\strut$\occconfig{{n^{(i-1)}}}$};
\node[above right] at (\tauvier,\nvier) {\strut$\occconfig{n^{(i+1)}}$};
\node [below right] at (0,\yrange) {b)};
\end{tikzpicture}}\\
  \mbox{\begin{tikzpicture}[semithick]
\newcommand{\xrange}{5}
\newcommand{\yrange}{3}
\newcommand{\M}{20}
\newcommand{\tauzwei}{\xrange*0.12}
\newcommand{\taudrei}{\xrange*0.3}
\newcommand{\tauvier}{\xrange*0.75}
\newcommand{\taufuenf}{\xrange*0.9}
\newcommand{\taua}{\xrange*0.5}
\newcommand{\taub}{\xrange*0.68}
\newcommand{\nzwei}{\yrange*0.2}
\newcommand{\ndrei}{\yrange*0.6}
\newcommand{\nvier}{\yrange*0.4}
\newcommand{\nnew}{\yrange*0.8}
\draw[->] (0,0) -- ++(\xrange+0.5*\xrange/\M,0) coordinate (xlabel) ;
\draw[->] (0,0) -- +(0,\yrange) coordinate (ylabel)  ;
\draw (\tauzwei,\nzwei) -- (\taudrei,\nzwei) --(\taudrei,\ndrei) --(\tauvier,\ndrei) -- (\tauvier,\nvier) -- (\taufuenf, \nvier) ;
\draw[very thick] (\taua,\ndrei) --(\taub,\ndrei);
\draw[dotted] (\tauzwei,\nzwei) -- +(-0.5,0);
\draw[dotted] (\taufuenf,\nvier) -- + (0.5,0);
\draw[dashed] (\taua,\ndrei) -- (\taua,\nnew) -- (\taub,\nnew) -- (\taub,\ndrei);
\node[below] at (\taudrei,0) {\strut$\tau_i$};
\node[below] at (\tauvier+0.1,0) {\strut$\tau_{i+1}$};
\node[below] at (\taua,0) {\strut$\tau_{a}$};
\node[below] at (\taub,0) {\strut$\tau_{b}$};
\draw (\taudrei,-0.1) -- (\taudrei,0.1);
\draw (\tauvier,-0.1) -- (\tauvier,0.1);
\draw (\taua,-0.1) -- (\taua,0.1);
\draw (\taub,-0.1) -- (\taub,0.1);
\node[above right] at (\taudrei,\ndrei) {\strut$\occconfig{n^{(i)}}$};
\node[above right] at (\taua-0.1,\nnew) {\strut$\occconfig{{n^{(i)}}\vphantom{n}'}$};
\node[above right] at (\tauvier,\nvier) {\strut$\occconfig{n^{(i+1)}}$};
\node[rotate=90] at (-0.5,0.5*\yrange) {Zustand $\occconfig{n}(\tau)$};
\node at (0.5*\xrange,-1.0) {Imaginäre Zeit $\tau$};
\node [below right] at (0,\yrange) {c)};
\end{tikzpicture}\begin{tikzpicture}[semithick]
\newcommand{\xrange}{5}
\newcommand{\yrange}{3}
\newcommand{\M}{20}
\newcommand{\tauzwei}{\xrange*0.12}
\newcommand{\taudrei}{\xrange*0.3}
\newcommand{\tauvier}{\xrange*0.75}
\newcommand{\taufuenf}{\xrange*0.9}
\newcommand{\taua}{\xrange*0.52}
\newcommand{\taub}{\taudrei}
\newcommand{\nzwei}{\yrange*0.2}
\newcommand{\ndrei}{\yrange*0.6}
\newcommand{\nvier}{\yrange*0.4}
\newcommand{\nnew}{\yrange*0.8}
\draw[->] (0,0) -- ++(\xrange+0.5*\xrange/\M,0) coordinate (xlabel) ;
\draw[->] (0,0) -- +(0,\yrange) coordinate (ylabel)  ;
\draw (\tauzwei,\nzwei) -- (\taudrei,\nzwei) --(\taudrei,\ndrei) --(\tauvier,\ndrei) -- (\tauvier,\nvier) -- (\taufuenf, \nvier) ;
\draw[very thick] (\taudrei,\ndrei) -- (\taua,\ndrei);
\draw[dotted] (\tauzwei,\nzwei) -- +(-0.5,0);
\draw[dotted] (\taufuenf,\nvier) -- + (0.5,0);
\draw[dashed] (\taua,\ndrei) -- (\taua,\nnew) -- (\taub,\nnew) -- (\taub,\ndrei);
\node[below] at (\taudrei,0) {\strut$\tau_i$};
\node[below] at (\tauvier,0) {\strut$\tau_{i+1}$};
\node[below] at (\taua,0) {\strut$\tau_{a}$};
\draw (\taudrei,-0.1) -- (\taudrei,0.1);
\draw (\tauvier,-0.1) -- (\tauvier,0.1);
\draw (\taua,-0.1) -- (\taua,0.1);
\node[above right] at (\taua,\ndrei) {\strut$\occconfig{n^{(i)}}$};
\node[above right] at (\taudrei,\nnew) {\strut$\occconfig{{n^{(i)}}\vphantom{n}'}$};
\node[above right] at (\tauvier,\nvier) {\strut$\occconfig{n^{(i+1)}}$};
\node at (0.5*\xrange,-1.0) {Imaginäre Zeit $\tau$};
\node [below right] at (0,\yrange) {d)};
\end{tikzpicture}}
  \caption{Visualization of the Monte Carlo steps. All pictures only show the relevant part of the path in continuous imaginary time. Dashed lines indicate the new path after the step has been performed. Thick lines mark the part of the path that will be removed. a) Step 1: Change state (determinant). The selected state is at $\ket{\occconfig{n^{(i)}}}$. b) Step 2: Shift kink. The selected kink is at $\tau_i$. Possible new kink positions are in between the kinks $\tau_{i-1}$ (not shown) and $\tau_{i+1}$. c) Step 3: Add pair of kinks. The selected interval is $\brackets{\tau_i,\tau_{i+1}}$. d) Step 4: Add single kink. The added kink is at $\tau_a$, and the selected interval is $\brackets{\tau_i,\tau_{i+1}}$.}
  \label{fig:MonteCarloSchritte}
\end{figure} 

\begin{enumerate}
  \item  Change state:
  \begin{enumerate}[i.]
    \item Randomly select one of the $N_S$ states $\ket{\occconfig{n^{(i)}}}$ of the path. For a configuration without kinks $(K=0)$ it is $N_S=1$, else $N_S=K$.
    \item Select randomly either a) a one-particle or b) a two-particle excitation
    \begin{enumerate}[a)]
      \item Randomly select one of the $N$ occupied and one of the $N_B-N$ unoccupied one-particle orbitals. Invert the occupation of both orbitals.
      \item Randomly select two of the $N$ occupied and two of the $N_B-N$ unoccupied one-particle orbitals. Invert the occupation of selected orbitals.
    \end{enumerate}
    \item If $K\geq 2$, check the following two cases:
       \begin{enumerate}[A)]
         \item If for a new many-particle state $\ket{\occconfig{n^{(i)\prime}}}$ $\ket{\occconfig{n^{(i)\prime}}}=\ket{\occconfig{n^{(i-1)}}}$ \underline{or} $\ket{\occconfig{n^{(i)\prime}}}=\ket{\occconfig{n^{(i+1)}}}$\par
         $\Longrightarrow$ Remove the corresponding kink.
         \item If for a new many-particle state $\ket{\occconfig{n^{(i)\prime}}}$ $\ket{\occconfig{n^{(i)\prime}}}=\ket{\occconfig{n^{(i-1)}}}$ \underline{and} $\ket{\occconfig{n^{(i)\prime}}}=\ket{\occconfig{n^{(i+1)}}}$\par
         $\Longrightarrow$ Remove both kinks.
        \end{enumerate}
        If none of the above cases applies or if $K=0$, then  
        \begin{enumerate}[A)]\setcounter{enumiii}{2}
  \item the number of kinks remains unchanged. 
\end{enumerate}
  \end{enumerate}
 
 \item Shift kink
    \begin{enumerate}[i.]
      \item Randomly select one of the $K$ kinks with time $\tau_i$.
      \item Randomly select a new kink position $\tau_i^\prime$ from the interval $[\tau_{i-1},\tau_{i+1}]$, where $\tau_{i-1}$ and $\tau_{i+1}$ are the times of the adjacent kinks. Hence, the ordering of the kinks remains unchanged. 
    \end{enumerate}

 \item Add pair of kinks
    \begin{enumerate}[i.]
      \item Randomly select one of the $K+1$ intervals $[\tau_i,\tau_{i+1}]$.
      \item Randomly select two new kink positions within the interval $\tau_a<\tau_b\in[\tau_i,\tau_{i+1}]$.
      \item Change the state on the interval $[\tau_a,\tau_b]$ as for MC move 1:
       \par Select randomly either a) a one-particle or b) a two-particle excitation
    \begin{enumerate}[a)]
      \item Randomly select one of the $N$ occupied and one of the $N_B-N$ unoccupied one-particle orbitals. Invert the occupation of both orbitals.
      \item Randomly select two of the $N$ occupied and two of the $N_B-N$ unoccupied one-particle orbitals. Invert the occupation of the selected orbitals.
    \end{enumerate}  
    \end{enumerate}
 \item Add a kink
     \begin{enumerate}[i.]
       \item Randomly select one of the $K$ kinks with time $\tau_i$.
       \item Randomly choose the interval $[\tau_{i-1},\tau_i]$  or $[\tau_i,\tau_{i+1}]$ for the new kink to be added. 
       \item Randomly select the new kink position $\tau_a$ in the chosen interval.
       \item Depending on the interval, change the state on the interval $[\tau_a,\tau_i]$ or $[\tau_i,\tau_a]$ as for step 1:
       \par Select randomly either a) a one-particle or b) a two particle excitation
    \begin{enumerate}[a)]
      \item Randomly select one of the $N$ occupied and one of the $N_B-N$ unoccupied one-particle orbitals. Invert the occupation of both orbitals.
      \item Randomly select two of the $N$ occupied and two of the $N_B-N$ unoccupied one-particle orbitals. Invert the occupation of selected orbitals.
    \end{enumerate}    
    \end{enumerate}
    \end{enumerate}

These four steps are selected with a kink-dependent probability. If $K=0$, only steps 1 and 3 are selected with the probabilities $p_{\text{cs}}$ and $p_{\text{add2}}$, respectively. In all other cases also the steps 2 and 4 are possible with the probabilities $p_{\text{sk}}$  and $p_{\text{add1}}$. Then, the probabilities $p_{\text{cs}}$ and $p_{\text{add2}}$ are adjusted. In the implementation which was used to produce the results of Sec.~\ref{sec:results} all currently possible steps are equally selected. 
\subsection{Acceptance probabilities of the Monte Carlo steps}
When determining the acceptance probabilities from the \emph{detailed balance} 
(cf. Eq. (\ref{detailed_balance}) and (\ref{Akz})), one only has to solve the \emph{detailed balance} for pairs of Monte Carlo steps that yield the original configuration when performed consecutively. In addition, the weight (\ref{weight}) divided by the partition function is a true probability only with respect to the number of kinks and \emph{Slater determinants}, but with respect to the kink times, it is a probability density. Only after multiplying with the infinitesimal kink times $d\tau_i$, the latter becomes a true probability with respect to the kink times. For steps that change the number of kinks, this would result in an infinitely large or vanishing acceptance probability, as some of the $d\tau_i$ do not cancel in the fraction of the weights. Fortunately, it was shown that one can simply treat the weights divided by the partition function as probabilities with respect to the kink times. This leads to finite acceptance probabilities while still being mathematically correct, see e.g. \cite{prokofev_trieste_2008}. \par
Instead of giving the detailed balance for all steps, we will rather explain how to set up and the solve the \emph{detailed balance} for one specific step, and simplify the obtained result of the acceptance probability so that it can actually be implemented. We choose step 3 of Sec. \ref{steps} (add pair of kinks). The \emph{detailed balance} for this step takes the following form:
\begin{align}
&p_\text{add2}(C)\frac{1}{K+1}\frac{2}{(\tau_{i+1}-\tau_i)^2}E |W(C)| A(C\to C')
= p_\text{cs}(C')\frac{1}{K'}E |W(C')| A(C'\to C)\label{db}\;,
\end{align}     
where the weights $W(C)$ are defined by Eq. (\ref{weight}) and $A(C\to C')$ is the acceptance probability of the step that converts the configuration $C$ with $K$ kinks into the configuration $C'$ with $K'=K+2$ kinks. The sampling probability to add two specific kinks consists of the product of the probabilities to
\begin{itemize}
  \item select step 3 ($p_{\text{add}}=\frac{1}{2}$, if $K=0$ and $\frac{1}{4}$ else): $p_{\text{add2}}$\;. 
  \item select one of the $K+1$ intervals of the configuration $C$, where the two kinks shall be added: $\frac{1}{K+1}$\;.
  \item select two new kink times $\tau_a$ and $\tau_b$ from the chosen interval $[\tau_i,\tau_{i+1}]$. Since there are two possibilities of the time ordering of the two randomly selected times, and since we will name the smaller one $\tau_a$, we have to add an additional factor 2: $\frac{2}{(\tau_{i+1}-\tau_i)^2}$
  \item  choose one of the two possible changes of the determinant $\ket{\occconfig{n^{(i)}}}$: $E$. For one-particle excitation it is
  \begin{align}
  E=E_1=\frac{1}{2}\frac{1}{N}\frac{1}{(N_B-N)}\;,
  \end{align}
  where the factor $\frac{1}{2}$ is the probability to choose the one-particle excitation and $\frac{1}{N}\frac{1}{(N_B-N)}$, the probability to select one of the occupied and one of the unoccupied orbitals. Analogously, for the two-particle excitations, it is:
  \begin{align}
  E=E_2=\frac{1}{2}\frac{1}{N}\frac{1}{(N-1)}\frac{1}{(N_B-N)}\frac{1}{(N_B-N-1)}\;.
  \end{align}
\end{itemize}
The sampling probability to propose exactly that change which transfers the configuration $C'$ back to the configuration $C$ consists of the product of the probabilities to
 \begin{itemize}
   \item select step 1 (change state): $p_{\text{cs}}$, where $p_{\text{cs}}=\frac{1}{2}$ if $K'=0$ and $\frac{1}{4}$ else. Since $K'=K+2\neq0$, it is $p_{\text{cs}}(C')\equiv\frac{1}{4}$. 
   \item select one of the $K'=K+2$ Slater determinants of the configuration $C'$:$\frac{1}{K'}$
   \item choose exactly that excitation which transfers $\ket{\occconfig{n^{(i)'}}}$ into $\ket{\occconfig{n^{(i)}}}$: $E$ 
 \end{itemize}
By inserting $K'=K+2$ and  $p_{\text{cs}(C')}=\frac{1}{4}$ into (\ref{db}), we obtain the acceptance probability for step 3:
 \begin{align}
A_{\text{add2}}(C\to C')=\min{\left[1, \frac{(K+1)}{(K+2)}\frac{(\tau_{i+1}-\tau_i)^2}{8p_{\text{add2}}(C)}\frac{|W(C')|}{|W(C)|}\right]}.
\end{align}
It is highly inefficient to calculate the whole fraction of weights for each Monte Carlo step, since this would very much increase the computation time. Therefore, the fraction of the weights has to be simplified as far as possible. For that purpose, we rewrite the weights as follows:
\begin{align}
W(C)=&(-1)^K e^{-\sum_{j=0,\atop j \neq i}^{K} D_{\{n^{(j)}\}} (\tau_{j+1}-\tau_j)}e^{D_{\{n^{(i)}\}} (\tau_{i+1}-\tau_i)} \prod_{j=0}^{K-1} 
Y_{\{n^{(j)}\},\{n^{(j+1)}\}}\,,\nonumber\\
W(C')=&(-1)^{K{'}} e^{-\sum_{j=0,\atop j \neq i}^{K} D_{\{n^{(j)}\}} (\tau_{j+1}-\tau_j)}e^{D_{\{n^{(i)}\}} (\tau_{a}-\tau_i)} e^{D_{\{n^{({i'})}\}} (\tau_{b}-\tau_a)}e^{D_{\{n^{(i)}\}} (\tau_{i+1}-\tau_b)} \nonumber\\
&\times\prod_{j=0}^{K-1} 
Y_{\{n^{(j)}\},\{n^{(j+1)}\}}\times Y_{\{n^{(i)}\},\{n^{({i'})}\}}^2\;.
\end{align}
When keeping in mind Fig. \ref{fig:MonteCarloSchritte}c), rewriting it in this particular way seems reasonable. Now most of the factors cancel in the weight fraction and we end up with 
\begin{align}
A(C\to C')=\min{\left[1, \frac{(K+1)}{(K+2)}\frac{(\tau_{i+1}-\tau_i)^2}{8p_{\text{add2}}(C)}e^{-\Bigl(\bigl(D_{\{n^{(i)}\}}-D_{\{n^{({i'})}\}}\bigr) (\tau_{b}-\tau_a)\Bigr)}\right]}\;.
\label{acc_prob}
\end{align}
This is the implemented acceptance probability to add a kink pair in the interval $[\tau_i,\tau_{i+1}]$ at the times $\tau_a$ and $\tau_b$. The acceptance probabilities for the remaining three steps can be derived with similar considerations.  

\section{Results for Interacting Fermions in a One-Dimensional Harmonic Trap\label{sec:results}}

As an example for the applicability of the introduced \emph{Configuration Path Integral Monte Carlo} (CPIMC) algorithm, we calculate thermodynamic properties of a one-dimensional system of spin-polarized, Coulomb-interacting fermions in a harmonic trap\index{potential!harmonic|see{harmonic trap}}\index{harmonic trap}. Despite its simplicity, this system is general enough to draw conclusions about the behavior of the method when simulating more complex systems. In particular, the Hamiltonian contains a pair interaction of the most general form  $\op{W}=\sum_{ijkl}w_{ijkl}^-\ladderup_i\ladderup_j\ladderdown_l\ladderdown_k$. In contrast to earlier applications of the \emph{Continuous Time Path Integral Monte Carlo} method, which usually consider lattice models, we investigate a spatially continuous system. At first, the test system will be characterized, and the used units will be introduced. In the following section, results for expectation values of several thermodynamic observables will be presented and compared to results obtained from the exact \emph{Configuration Interaction} (CI) method, see e.g. \cite{helgaker_molecular_2000}, and from the \emph{Direct Path Integral Monte Carlo} (DPIMC) method~\cite{bonitz_introduction_2006}, respectively. Finally, the dependence of the efficiency of the CPIMC and DPIMC methods on different parameters will be investigated.

\subsection{The System}
The Hamiltonian of the considered system in natural units is given by
\begin{align}
\Ham = \sum_{\alpha=1}^N \Big(\frac{\op{p}_\alpha^2}{2m} + \frac{m\omega^2\op{r}^2_\alpha}{2} \Big) + \sum_{1\leq\alpha<\beta}^N \frac{e^2}{\sqrt{\abs{\op{r}_\alpha-\op{r}_\beta+\kappa^2}}}.
\end{align}
Here, $m$ is the mass of the fermions and $\omega$ the frequency of the trap. The parameter $\kappa$ is necessary in one dimension to prevent the divergence of the two-particle integrals of the Coulomb-interaction, cf. (\refeq{eq:Zweiteilchenintegrale}). Introducing the coupling parameter\index{coupling parameter!quantum} $\lambda={E_\text{C}}/{\omega}$ which denotes the ratio of the Coulomb-energy $E_\text{C}={e^2}/{r_0}$ with $r_0^2={1}/{(m\omega)}$ and the potential energy of the trap $\omega$, the Hamiltonian can be written in dimensionless units:
\begin{align}
\Ham = \frac{1}{2}\sum_{\alpha=1}^N (\op{p}_\alpha^2+\op{r}^2_\alpha ) + \sum_{1\leq\alpha<\beta}^N \frac{\lambda}{\sqrt{\abs{\op{r}_\alpha-\op{r}_\beta+\kappa^2}}}.
\end{align}
In these units, the energy is given by ${E}/{\omega}$, the chemical potential by ${\mu}/{\omega}$ and the temperature by ${T}/{\omega}$ and $\beta={\omega}/{T}$, respectively. 
\par
The eigenfunctions of the one-particle Hamiltonian are the well known eigenfunctions of the quantum mechanical harmonic oscillator. In this basis, the one- and two-particle integrals can be calculated in advance with high precision\footnote{The one- and two-particle integrals used in these calculations have been calculated by K. Balzer with Mathematica \cite{mathematica}.}. Furthermore the one- and two-particle integrals in the Hartree-Fock (HF) basis will be used in the following. They have been calculated for each specific value of the temperature, the particle number and the coupling strength\footnote{These integrals have been calculated with a program by K. Balzer.}.
\par
The results in this work have been obtained under the additional assumption of spin-polarized fermions, i.e. all particles have the same spin projection remaining fixed in the simulation. Such a system is realized e.g. in the presence of a strong magnetic field. It should be noted that this was chosen for simplicity and is not a restriction of the method at hand. Generalizations to systems with arbitrary spin configurations or even with a spin dependent Hamiltonian are straightforward.

\begin{figure}[tb]
\centering
  \includegraphics[width=\textwidth]{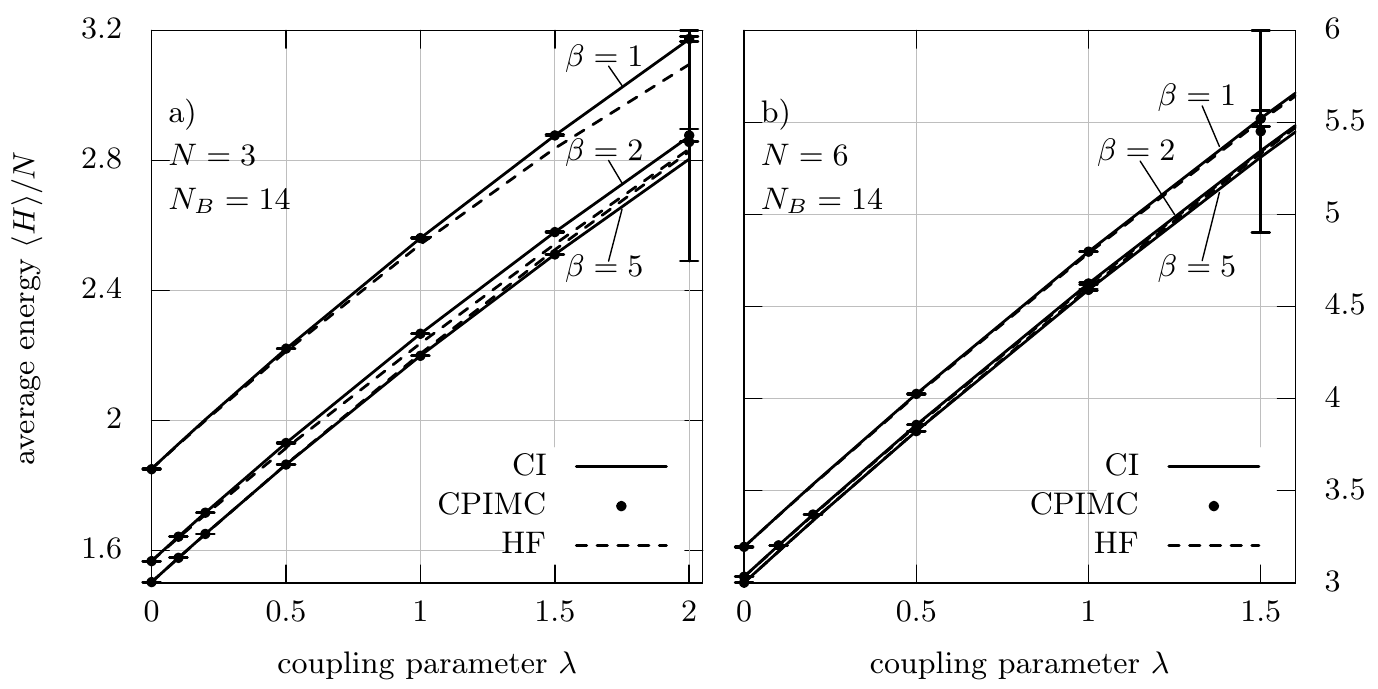}
  \caption[Average energy per particle versus coupling parameter.]{
  Average energy per particle versus coupling parameter for a) $N=3$ and b) $N=6$ particles and different temperatures. Comparison of CPIMC results (dots), CI results (solid lines) and HF results (dashed lines). Temperatures are given by labels and small arrows within the figure.
  Error bars show the statistical error of the CPIMC simulation.}
  \label{EnergyCPIMCoverLambda}
\end{figure}

\subsection{Numerical Results}

\begin{figure}[tb]
\centering
  \includegraphics[width=\textwidth]{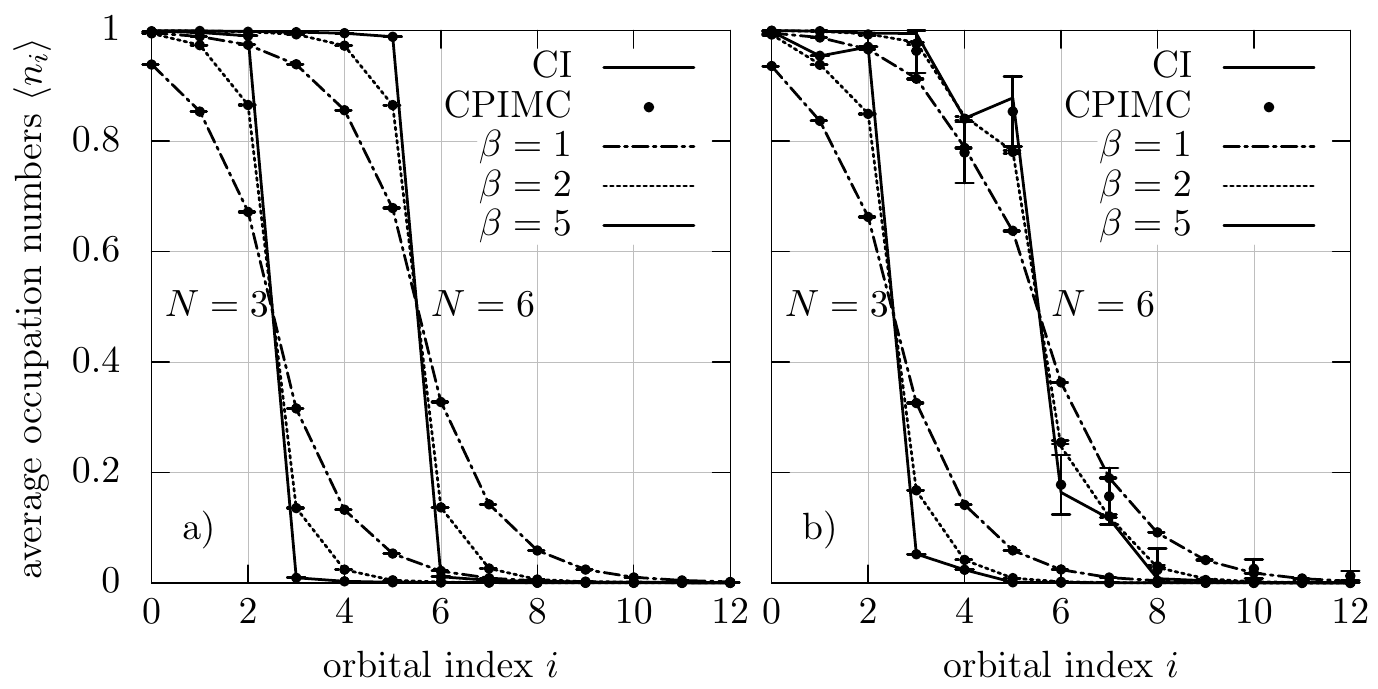}
  \caption[Average occupation numbers for different termperatures and particle numerbers.]{
  Average occupation numbers $\braket{\op{n}_i}$ of the one-particle orbitals $\ket{i}$. a) in the HF basis and b) in the basis of eigenfunctions of the ideal Hamiltonian. Shown are results obtained from CPIMC calculations (points with error bars) and CI results (lines) for the coupling strength $\lambda=1$ and the temperatures $\beta=1$ (dashed dotted lines), $\beta=2$ (dotted lines) and $\beta=5$ (solid lines). Adapted from \cite{schoof_configuration_2011}.}
  \label{BesetzungszahlenCPIMCvonT}
\end{figure}
 The results in this section were obtained with the CPIMC method using \num{5e9} Monte Carlo steps per data point, if not noted otherwise. Samples for the thermodynamic averages were taken every \num{5000} steps. This large number of steps is necessary because of the large auto-correlation time. In total, this results in \num{1e6} samples per observable.  The equilibration time was chosen to \num{5e7} steps. The runtimes for these calculations were in between \SI{3}{h} and \SI{12}{h} on a single-core CPU depending on the particle number, temperature and coupling parameter. 
 \par
First, we will consider the energy per particle, for it is a central and simple observable. Fig.~\ref{EnergyCPIMCoverLambda} shows the results for $N=3$ and $N=6$ particles obtained with 3 different methods: CPIMC, HF and CI\footnote{The finite temperature HF and CI programs were written by D. Hochstuhl.}. For HF and CI, the first $N_B=14$ eigenfunctions of the ideal one-particle Hamiltonian were used as basis functions. For CPIMC, the basis functions from the HF calculation were used. The CI results can be considered as reference data, because the only approximation of CI, the usage a finite basis set, is the same for all methods. For all shown coupling strengths and temperatures, the energies from the CI calculation match the energies from the MC calculation within the statistical errors\footnote{Shown is the standard error, see~(\refeq{error_root}). Within two times the standard error the results can be considered to coincide.}. For weak and moderate coupling, $\lambda\leq1$, the statistical error is small and the agreement is very good. For stronger couplings, the statistical error strongly depends on the temperature and on the particle number. While the precision for $N=3$ and $\beta=2$ is high enough to distinguish the CPIMC result from the HF result up to $\lambda=2$, this is not the case for $\beta=5$ or $N=6$. For $\lambda=\num{1.5}$, $N=6$ and $\beta=5$, the relative error exceeds \SI{100}{\%} and the corresponding data point is not shown.
\par
A quantity that is much more sensitive to systematical errors is the average occupation number $\braket{\op{n}_i}$ of each one-particle orbital $\ket{i}$. The occupation numbers of the orbitals in the HF and the ideal basis for different temperatures and particle numbers are shown in Fig.~\ref{BesetzungszahlenCPIMCvonT}. In the case of HF orbitals the shape of the curves is similar to the Fermi-Dirac distribution of non-interacting fermions in the grand-canonical ensemble. It should be noted that it is not possible to relate a specific energy to each orbital as in the case of basis functions that diagonalize the Hamiltonian, e.g. for ideal systems. Moreover the orbital energies of the HF orbitals have a meaning only within the HF approximation. Additionally, the HF orbitals with the same index differ from each other for different values of the temperature. For small temperatures, the shape of the curves approaches a step functions, but as long as the exact ground state does not equal the HF ground state (which is the general case due to the interaction), there will be non-zero occupation numbers above the step even in the ground state at $T=0$. There is an excellent agreement between the CPIMC and CI results for all shown parameters.  
\par
Considering the results for the occupation numbers of the ideal orbitals, the non-monotonic behavior is obvious. This behavior is confirmed by the CI results, which agree well within the statistical error. However, the statistical errors in these calculations are much larger than in the calculations using the HF orbitals, which will be discussed later.
\par
\begin{figure}[tb]
\centering
  \includegraphics[width=\textwidth]{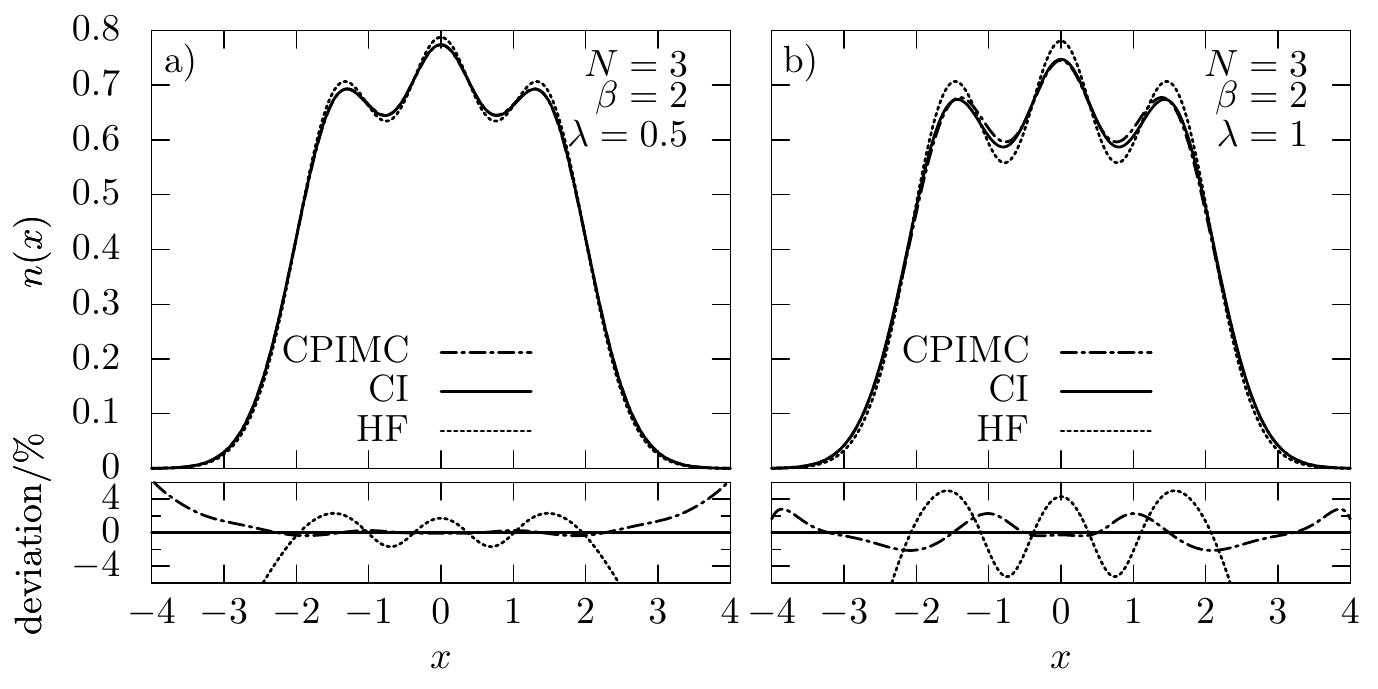}
  \caption[ One-particle density for $N=3$ particles, $\beta=2$ and different coupling strength.]{
  One-particle density $n(x)$ for $N=3$ particles, $\beta=2$ and a) $\lambda=\num{0.5}$ and b) $\lambda=\num{1}$. Solid lines indicate CI results, which are considered to be exact. CPIMC and HF results are indicated by dashed dotted and dotted lines, respectively. Below, the relative deviation $(n(x)-n_\text{CI}(x))/n_\text{CI}(x)$ is shown.}
  \label{VergleichBesetzungszahlenN3}
\end{figure}
The final observable, for which we will compare the results between the different methods in this chapter, is the one-particle density $n(x)$ where $x$ denotes the spatial coordinate. As can be seen from Eq.~(\refeq{eq:Einteilchendichte}), all matrix elements of the one-particle density matrix contribute to this quantity. Compared to the diagonal elements  $d_{ii}=\braket{\op{n}_i}$ shown before, the off-diagonal elements $d_{ij}$ with $i\neq j$ converge much slower. The one-particle density contains a large statistical error since each of the off-diagonal elements can be determined only in certain configurations of the Markov-chain, see (\refeq{eq:KinkEinteilchendichtematrix}). Correspondingly, a larger number of MC steps would be necessary for a precise calculation of the whole one-particle density matrix. Nevertheless, Fig.~\ref{VergleichBesetzungszahlenN3} shows that for $N=3$ particles, a temperature of $\beta=2$, and a coupling strength of $\lambda=\num{0.5}$ and $\lambda=\num{1}$ the agreement of the CPIMC results with the exact density is better than in the case of the HF results. The relative deviation amounts to only a few per cent. This can be explained by the diagonal elements (which can be calculated efficiently, see Fig.~\ref{BesetzungszahlenCPIMCvonT}) dominating the contribution to the density. Additionally, the statistical errors in the off-diagonal elements may partly cancel each other in the calculation of the one-particle density.
\par
For all thermodynamic quantities presented in this chapter, we could demonstrate the very good agreement within the statistical error between the results obtained from the CPIMC method and the reference results obtained from the exact diagonalization of the Hamiltonian. For weak and moderate coupling strengths $\lambda\leq1$ and moderate temperatures, i.e. $\beta=1$ and $\beta=2$, results can be obtained with high precision within several hours of runtime on a single CPU. 

\subsection{Fermion sign problem}
\begin{figure}[tb]
\centering
  \includegraphics[width=\textwidth]{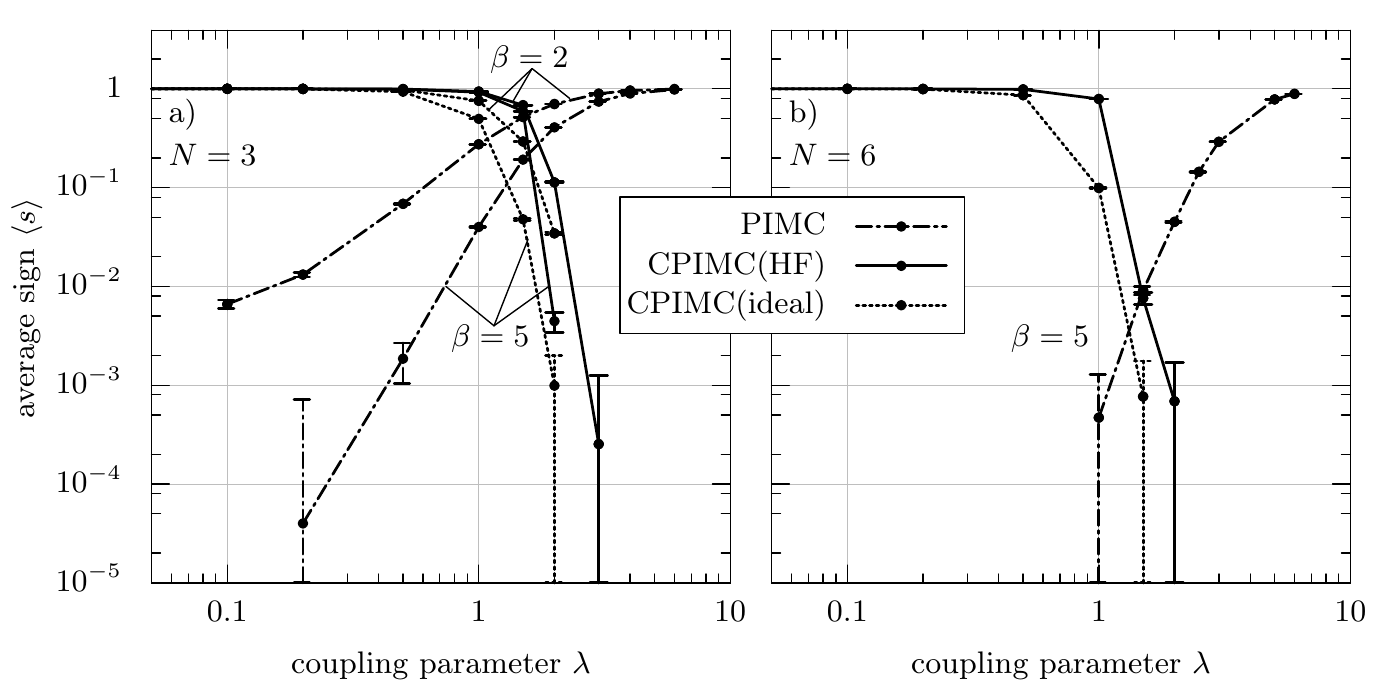}
  \caption[Average sign versus coupling parameter.]{
  Average sign versus coupling parameter for different temperatures and a) $N=3$ and b) $N=6$ particles. Shown are results obtained from CPIMC calculations with the HF basis (solid lines) and with the ideal basis (dotted lines). Dashed dotted lines indicate results from DPIMC calculations. Temperatures are given by labels and small arrows within the figure. Error bars show the statistical errors. Note the log scale. Adapted from \cite{schoof_configuration_2011}.}
  \label{MittleresVorzeichenVonNNB14}
\end{figure}
In the last section, the focus lay on the exact calculation of thermodynamic averages of physical observables. In this section, we will investigate the main factor that determines the efficiency of QMC methods: the fermion sign problem. Taking the modulus of the weights to allow for the MC sampling of configurations with negative weights introduces the average sign $\braket{s}'$ in Eq.~(\refeq{sign}). For small average sign, the statistical error of MC calculations is approximately inverse proportional to it. For path integral Monte Carlo methods in coordinate space, e.g. DPIMC, the average sign decreases exponentially with the particle number $N$ and the inverse temperature $\beta$. This leads to exponentially growing statistical errors. Alternatively, if results with a fixed precision are targeted, this leads to exponentially growing runtimes, see Eq.~(\refeq{eq:mc_error}).
\par
It turns out that the dependence of the average sign on temperature and particle number for CPIMC is essentially the same as for DPIMC, namely exponential.  This is as expected since the general behavior is governed by Eq.~(\refeq{eq:mc_error}). The dependence of the average sign on the coupling strength is shown in Fig.~\ref{MittleresVorzeichenVonNNB14} for $N=3$ and $N=6$ particles and different temperatures. Additionally, the figure shows the average sign of DPIMC calculations for the same parameters\footnote{The data was produced by A. Filinov.}. For large coupling strength, $\lambda\gg1$, the average sign of DPIMC is close to $\num{1}$ and therefore, calculations with DPIMC are highly efficient. For decreasing $\lambda$, the sign decreases and calculations with DPIMC become more demanding. The simulation of the ideal system with $\lambda=0$ is infeasible. The reason for this behavior is the following: For strong repulsive interactions, the overlap of the wave functions of the particles is small and the statistics resemble those of a classical system of point-like particles. The weaker the Coulomb repulsion compared to the strength of the trap, the larger is the overlap of the wave functions. With larger overlap, particle exchange, which is the source of the alternating sign in DPIMC, becomes more important.
\par
CPIMC calculations, on the other hand, show a behavior completely opposite to that of DPIMC. While for the ideal system there exists no sign problem at all, the average sign decreases with increasing coupling parameter. This decrease becomes very fast at $\lambda\approx1$, and calculations for larger coupling parameters become inefficient. The value of $\lambda$ for which this rapid decrease of the average sign starts depends on the temperature and the particle number. For lower temperature and larger particle number, the average sign decreases earlier. Fig.~\ref{MittleresVorzeichenVonNNB14} also shows the influence of the choice of the basis on the average sign. Using HF basis functions, the average sign and therefore the efficiency of CPIMC can be increased by several magnitudes.

\section{Summary and Outlook}
In this chapter, we presented a path integral Monte Carlo method in the Slater determinant space for the exact calculation of thermodynamic properties of correlated fermionic many-particle systems. We introduced the appropriate description of the physical system in terms of second quantization and described the transition from the coordinate representation to the occupation number representation of the path integral in detail. In contrast to the widely used application of the underlying formulation to lattice models with onsite (or short-range) interaction, such as the Hubbard model, the method at hand allows for ab initio calculations of spatially continuous systems with arbitrary interaction. We gave a detailed description of the Monte Carlo steps including the acceptance probabilities for the exmple of one step. At the end we demonstrated the application of this method to a test system of Coulomb-interacting fermions in a one-dimensional harmonic trap. While we have presented only a small selection of thermodynamic quantities, our simulations are able to yield all other thermodynamic quantities. Moreover, it was recently demonstrated that also dynamic quantities, including the energy spectrum and collective excitation spectrum, can be efficiently computed within quantum Monte Carlo simulations \cite{afilinov_pra12} 
\par
For the test system, the comparison of the results  between the CPIMC and DPIMC method revealed a very interesting complementary behavior of the average sign, depending on the coupling strength. While the calculation of thermodynamic expectation values with DPIMC is possible for strong coupling and low degeneracy, CPIMC allows to calculate these quantities for weakly to moderately coupled and highly degenerate many-body systems, which are not accessible with DPIMC. Therefore the presented method helps to extend the parameter range for which \emph{ab initio} QMC simulations are feasible. Furthermore, as mentioned in the introduction, strongly degenerate weakly coupled electrons are of key importance for many modern plasma applications, so here CPIMC is expected to be very valuable.
\par
Despite the promising results shown in this chapter, further work has to be done to improve the efficiency of the method. The exponential dependence of the average sign on the temperature and particle number still exists, which opens an increasing gap in the ranges of coupling parameters for which either CPIMC or DPIMC methods are efficient. This gap can, at the moment, be bridged only by additional approximations. Approaches to this problem can be assigned to two classes. The first class contains improvements to the algorithm that increase the effective number of MC samples per time, e.g. by optimizing the acceptance probability of the steps and by decreasing the auto-correlation time. The so-called heat bath idea is such an approach. By choosing the sampling probability to be proportional to the transition probability, the acceptance probability becomes unity. This is possible if the normalization for the transition can be calculated efficiently, e.g. for choosing the time of the kinks in the MC steps. Another promising approach in this class is the so-called worm algorithm \cite{prokofev_exact_1998} which also allows for more efficient sampling of off-diagonal quantities related to the Matsubara Green function, including the one-particle density. Other approaches aim at reducing the sign problem directly. One method in this class is the so-called blocking, which tries to analytically combine configurations of opposite sign into groups with a much lower modulus of the weight than the single configurations. 
\par
To achieve even higher efficiency additional approximations can be used. The similarities of the CPIMC method to MC on a lattice (such as the Hubbard model) as well as to CI methods, give access to a large variety of approximations developed for these latter methods. As an example we mention the restricted active space (RAS) approach well known in quantum chemistry, see \cite{TDRASCI}. Special cases are the  ``CI-Singles'' and ``CI-Doubles'' approximations. The main idea is to restrict the occupation number of the orbitals in different energy (or spatial) regions. For example, one can require that the $n$ orbitals within the lowest energy interval must be occupied by at least $n-1$ particles, the $m$ orbitals within the next-higher energy interval be occupied by at least $m-2$ particles and so on, until the orbitals with the highest energy are occupied at most by one particle. These restrictions are arbitrary but can be motivated by physical insights. This approach strongly decreases the size of the configuration space and consequently the sign problem. The advantage of this approximation is that it is easy to control and to fine-tune. Calculations with gradually reduced restrictions allow one to estimate the introduced systematical error. With these improvements applications of CPIMC to larger particle numbers and, eventually, to macroscopic quantum plasmas seem to be possible.

\section*{Acknowledgements}
We are grateful to K. Balzer, A. Filinov and D. Hochstuhl for providing supplementary data shown in this chapter.
We acknowledge financial support by the Deutsche Forschungsgemeinschaft via grant BO1366-10 and a grant for CPU time at the HLRN.
\newpage

\end{document}